\documentclass[journal]{IEEEtran}
\usepackage{graphicx}
\usepackage{caption}
\usepackage{multirow}
\usepackage{color}

\usepackage{enumitem}

\newcommand\dtrain{\ensuremath{\mathcal{D}_{tr}}}
\newcommand\users{\ensuremath{\mathcal{U}}}
\newcommand\Hcal{\ensuremath{\mathcal{H}}}

\newcommand{\added}[1]{{\color{black}{#1}}}

\newcommand{\myparagraph}[1]{ \noindent \textbf{#1.}}
\newcommand{\ignore}[1]{}

\begin{document}
\title{The House That Knows You: \\ User Authentication Based on IoT Data}

%

%\titlerunning{Abbreviated paper title}

% If the paper title is too long for the running head, you can set

% an abbreviated paper title here

%

\author{Talha~Ongun,
	Oliver~Spohngellert,
	Alina~Oprea,
	Cristina~Nita-Rotaru,
	Mihai~Christodorescu,
	Negin~Salajegheh
\thanks{This work was supported by
	Visa Research. (Corresponding
	authors: Talha Ongun; Alina Oprea.)}%
\thanks{T. Ongun, O. Spohngellert, A. Oprea, C. Nita-Rotaru are with 
	Khoury College of Computer Sciences at Northeastern University, Boston, MA, 02115 USA (email: ongun.t@northeastern.edu; spohngellert.o@northeastern.edu; a.oprea@northeastern.edu; c.nitarotaru@northeastern.edu ) }%
\thanks{M. Christodorescu, and N. Salajegheh were with 
	Visa Research, Palo Alto, CA 94306, USA (email: mihai.christodorescu@visa.com ) }%
}

\maketitle              % typeset the header of the contribution

\begin{abstract}
	Home-based Internet of Things (IoT) devices have gained in popularity and many households have become ``smart'' by using  devices such as smart sensors, locks, and voice-based assistants.  Traditional authentication methods such as passwords, biometrics or multi-factor (using SMS or email) are either not applicable in the smart home setting, or they are inconvenient as they break the natural flow of interaction with these devices. Voice-based biometrics are limited due to safety and privacy concerns. Given the limitations of existing authentication techniques, we explore  new opportunities for user authentication in smart home environments. Specifically, we design a novel authentication method based on behavioral features extracted from user interactions with IoT devices. We perform an IRB-approved user study in the IoT lab at our university over a period of three weeks. We collect network traffic from multiple users interacting with 15 IoT devices in our lab and extract a large number of features to capture user activity. We experiment with multiple classification algorithms and also design an ensemble classifier with two models using disjoint set of features. We demonstrate that our ensemble model can classify five users with 97\% accuracy. The behavioral authentication modules could help address the new challenges emerging with smart home ecosystems and they open up the possibility of creating flexible policies for authorization and access control.

\end{abstract}

%\keywords{Behavioral authentication; Internet of Things; Machine learning}
\begin{IEEEkeywords}
	Behavioral authentication, machine learning, smart home
\end{IEEEkeywords}

\IEEEpeerreviewmaketitle

% !TEX root = iot_auth_main.tex

\section{Introduction}

User authentication is one of the major challenges in security, and while a diverse set of solutions
have been proposed, the quest for the \added{optimal} solution continues.
Traditional user authentication methods include  passwords, hardware tokens, and biometrics.  Given
the limitations of passwords~\cite{cranor2018pwd,cranor2016pwdmobile,cranor2016pwdperception}
several well-known techniques are used in practice to augment them,
such as multi-factor authentication based on either a hardware token or a different channel  (e.g. email, SMS).
These methods impede the usability of authentication, and introduce additional latency~\cite{siadati2017mind,de2013comparative}.
Biometrics-based authentication has been extensively used in various forms (facial recognition, fingerprints, or retina scans), but revocation and user impersonations are well-recognized limitations of these methods.

\ignore{The omnipresence of IoT devices, as well as the smart households that use smart devices such as
sensors, locks, and voice-based personal assistants, have only made the environment for authentication more
complex, bringing new challenges.}
The omnipresence of IoT devices  such as sensors, locks, and voice-based personal assistants has made authentication in smart households  more complex, bringing new challenges. IoT devices are fundamentally different from personal devices, and traditional authentication methods are either not applicable or break the natural flow of interaction with these devices. Deploying a screen or input pad for each of these devices is either not practical or costly. Some devices such as Amazon Echo Dot can perform voice-based biometric authentication as an optional feature. However, this feature can be circumvented as voices can be spoofed, and users might not be comfortable with their voices being recognized due to privacy concerns~\cite{PrivacyIoT}. Therefore, better authentication methods are needed in the context of smart homes. One of the opportunities in solving this problem in smart home environments is the relatively small number of users for which
authentication must be provided. Most American households range in size from two to six members. Multi-class classification algorithms can be trained for such small sets, and with minimal
effort, it can be ensured that adversaries do not interfere at training time.
However, there are several challenges in designing user authentication methods for
household IoT devices including: supporting a diverse set of devices with minimal
changes in their operation, maintaining user privacy,
and supporting a diverse set of policies.  Data that captures detailed user behavior (such as device logs) tend to be device-specific and more invasive
in terms of privacy.

Our approach addresses these challenges by using the
IoT devices' network traffic  from the home router as the data source. In order to preserve user privacy,
we collect and leverage the information in the headers of HTTPS packets for our authentication system. This data includes minimal information (timing, ports, bytes sent and received) with low risk to user privacy. Additionally, to protect user privacy we make the design choice of performing the model training and testing locally in the user's home, rather than remotely in the cloud. This implies that the authentication module needs to reside in the smart home. To address the mismatch between user-level semantics and network-level semantics, we
observe the network traffic for a continuous time window and design a machine learning (ML) model that uses features aggregated over a recent time window to predict the user in the room. Finally, we train a multi-class classifier to predict the likelihood that a certain user is in the room. The model is trained based on historical data obtained in controlled settings (knowing which user is in the room), and then used at testing time to classify users by generating in real time an authentication score based on the most recent observed activity.

In this paper, we propose for the first time behavioral authentication models in the smart home that authenticate users based on their flow of interactions with multiple IoT devices.
We leverage the IoT lab at our institution set up as a studio apartment, and purchase a set of 15 IoT devices from different categories, including voice assistants, smart kitchen appliances, and entertainment devices. We make use of the existing monitoring infrastructure in the lab to collect network packets (pcap files) from all these devices and design an IRB-approved user study with multiple users participating in data collection over a period of three weeks. During the training period, we use labeled data of user sessions and extract features from HTTPS headers that capture user interaction with the IoT devices over a continuous time window. We design and test several ML classification algorithms for predicting the likelihood that a certain user is in the room. During the regular user interaction with the IoT devices, the authentication module continuously receives data extracted from the most recent observation window and computes an authentication score for each user. The score is made available to upper-level authorization systems that can implement flexible policies according to the level of risk tolerated. For example, these policies can be used for local services such as logging into a laptop,  performing operations on IoT devices, or accessing cloud services, such as performing financial transactions.
Our experimental evaluation shows that a Gradient Boosting classifier achieves 81\% precision and 80\% recall at classifying six users, based on IoT device features. For five users, the precision and recall are both 92\% for a 25-minute observation window. We also design a high-confidence ensemble classifier with two models trained on disjoint sets of features (one on 420 device features and the second on 2910 domain features). The ensemble generates an authentication score only when the two models agree on the prediction. An ensemble of two Gradient Boosting models achieves an F1 score of 0.86 for six users and 0.97 for five users (at the cost of not generating scores in 16 out of 91 sessions).
The rest of the paper is organized as follows. We define our problem and adversarial model in Section \ref{sec:bk}.
We describe our system design in Section \ref{sec:system}, and our user study and data collection in Section \ref{sec:data}.
We present the evaluation in Section \ref{sec:evaluation} and related work in Section \ref{sec:relwork}.
We conclude the paper in Section \ref{sec:concl}.

% !TEX root = iot_auth_main.tex

\section{Preliminaries and Background}
\label{sec:bk}
In this section, we provide an overview of methods for user authentication, problem definition, and the adversarial model we consider in this work.

\subsection{User Authentication}

\ignore{
User authentication is a fundamental security process that has been studied extensively throughout the years. Traditional user authentication methods are based on \emph{what users know} (e.g. passwords), \emph{what they have} (e.g. hardware token), or \emph{what they are} (e.g. biometrics).  Even though extensive research  shows  password limitations~\cite{cranor2018pwd,cranor2016pwdmobile,cranor2016pwdperception}, passwords are still the de-facto standard for authentication in corporate and home environments. Users are shown to pick weak passwords and they tend to reuse them across different services \cite{worstpasswords}. Critical services do not always deploy proper password data protection techniques, resulting in increased risk for users in face of data breaches~\cite{facebookplain}. Several well-known techniques are used in practice to augment passwords. Multi-factor authentication methods use either a hardware token or a different channel  (e.g. email, SMS) for increasing the confidence of authenticating users. These methods impede the usability of authentication, and introduce additional latency~\cite{siadati2017mind,de2013comparative}.
Biometrics-based authentication such as keystroke dynamics, fingerprints, facial recognition, and voice recognition  provide a convenient and reliable way to identify users~\cite{wayman2005introduction,SurveyBiometrics,SurveyBiometricWearable}. However, revocation and user impersonation are well-known challenge for these methods, in addition to privacy implications for users~\cite{wolf2019pretty}.
}

Traditional user authentication methods are based on \emph{what users know} (e.g. passwords), \emph{what they have} (e.g. hardware token), or \emph{what they are} (e.g. biometrics).  Even though extensive research  shows  password limitations~\cite{cranor2018pwd,cranor2016pwdmobile,cranor2016pwdperception}, passwords are still the de-facto standard for authentication in corporate and home environments. Biometrics-based authentication such as keystroke dynamics, fingerprints, facial recognition, and voice recognition  provide a convenient and reliable way to identify users~\cite{wayman2005introduction,SurveyBiometrics,SurveyBiometricWearable}. However, revocation and user impersonation are well-known challenge for these methods, in addition to privacy implications for users~\cite{wolf2019pretty}. 
Behavior-based authentication  rely on models that learn the unique
way  users interact with devices like smartphones,
tablets or touch screens. 
In the context of mobile devices, several systems for continuous authentication have been designed~\cite{shi2010implicit,riva2012progressive}.  They leverage a number of behavior features from call logs, browser history, application usage, and location. User authentication based on WiFi signal of IoT devices has been also investigated recently~\cite{shi2017smart}. Multi-factor authentication methods use either a hardware token or a behavioral modelfor increasing the confidence of authenticating users, but they introduce  usability challenges~\cite{siadati2017mind,de2013comparative}.

\subsection{Problem Definition}

With the widespread usage of IoT devices in many households, new challenges and opportunities \added{are emerging}. We pose for the first time \emph{the problem of designing usable, behavioral-based authentication models based on users naturally interacting with IoT devices in their homes or familiar environments}. We are interested in designing a lightweight authentication module for smart homes that uses information extracted from monitoring the user interaction with IoT devices. The system should be able to identify with high accuracy each user from a fixed set of known users based on their behavioral characteristics.   With user privacy as the \added{principal} consideration in our design, we intend to leverage a minimum number of attributes that are universally applicable to a range of IoT devices from different manufacturers. We consider here mainly authentication based on behavioral features extracted from network traffic generated by IoT devices in a smart home. We envision \added{that} such a system would learn from historical observations and create user profiles  during a training phase. At testing time, the authentication module generates an authentication score mapped to the likelihood that a certain user is in the room based on the most recent observation of IoT device activity. The authentication module needs to observe users during a contiguous observation window to capture a sequence of user actions. The authentication scores can be used by higher-level applications to create flexible authorization policies that support various levels of risk.

\subsection{Adversarial Model} 

We consider two main classes of adversaries:

\myparagraph{Local adversaries} This adversary might be in close proximity or even inside the home environment, and interact with some of the IoT devices in the home. This model captures visitors, neighbors, kids, or roommates sharing the home. We would like to prevent such an adversary from performing actions like changing smart lock settings or making purchases on Alexa on behalf of real users. The premise is that a local adversary might perform shoulder-surfing and find out a user's PIN or even a weak password. However, mimicking the interaction of the user with IoT devices, particularly the sequence of actions, is inherently difficult. The adversary might learn how the user interacts with one device (e.g., the coffee brewer) and could replicate such behavior. The strength of our system comes from combining signals based on interaction with \emph{all devices in the smart home}, which we believe is much more difficult to impersonate.

\myparagraph{Remote adversaries} Remote adversaries might compromise user machines and credentials, and use existing network protocols and APIs to send traffic to the home router or directly to IoT devices. Their motivation is user impersonation to gain access to sensitive devices in the home or perform  harmful actions. Another possible threat is a malicious app installed on a user's smart phone that sends network traffic to interact with the IoT devices. The strongest type of remote adversary is one that leverages vulnerabilities in the IoT devices, and eventually obtains access to a limited set of IoT devices in the smart home. An attacker with full access to an IoT device can inject network traffic and simulate user actions of its choice. However, we assume that the attacker will not be able to compromise \emph{all the IoT devices at once} and gain complete access to the smart home. The resilience of our authentication system is enabled by the large set of features extracted from user interaction with multiple IoT devices over time.  
We envision that data collection and the authentication module reside in a trusted device, such as the home router or a server. Compromising the authentication system by physically accessing it is not within our scope. We also assume that adversaries are not in the system at training time. Poisoning attacks are training-time attacks against ML models~\cite{biggio2012poisoning,Xiao15,Steinhardt17,Jagielski18}, and protecting against them is orthogonal to our work. Potential mitigations against poisoning include the use of multi-factor authentication during training.

% !TEX root = iot_auth_main.tex

\section{System design}
\label{sec:system}

In this section, we describe the goals motivating the design of our system and then give an overview
of our approach.
	\vspace{-4mm}

\begin{figure*}[t]
	\centering
	\includegraphics[width=0.85\linewidth]{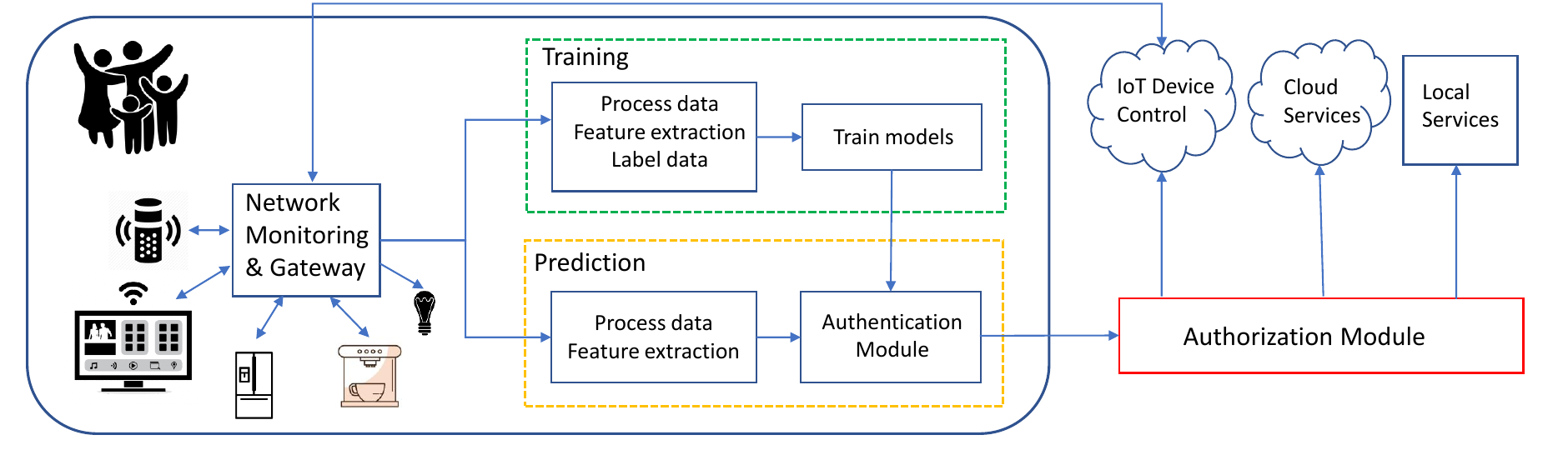}
				%\vspace{-3mm}
	\caption{Architecture diagram of the dataflow and proposed authentication process.}
	\label{fig:architecture}
			%\vspace{-6mm}
\end{figure*}

\subsection{Design Goals}

 In designing our system for user authentication in smart homes, we set forward the following design goals:

%\begin{enumerate}
\myparagraph{Operate in multi-user smart homes} We intend our system to be used in a smart home environment with several users actively interacting with  IoT devices. The average American household size in 2018 was 2.53, with most households (98.66\%) having a size of 6 or fewer. Our system should be able to classify users from a small set of previously-seen users with good accuracy.

\myparagraph{Support multiple heterogenous IoT devices} Compared to traditional computing platforms, IoT devices are extremely heterogeneous and diverse. This raises the challenge of designing a flexible  monitoring platform that  supports devices from different manufacturers and can be easily extended to include new devices. While device logs are specific to the device type, network traffic provides a universal data type for monitoring IoT devices connected to the Internet.

Available data usually consists of event logs and network traffic. Although some devices provide event logs through their cloud interface which can be collected in an automated way, not all devices give access to this data. We observe that network traffic is much more general and more likely to support the diversity of IoT devices as most of the IoT devices are connected to the internet by definition.

\myparagraph{Maintain user privacy} User privacy is an important design consideration, given the nature of the applications and
the environment where users perform their actions. We intend our system to work without access to user personal identifiable information (PII). Most network traffic of IoT devices is over HTTPS, with encrypted payloads in the transmitted packets.

\myparagraph{Capture user actions for behavioral modeling} User behavior is defined by their interaction with the devices and the
applications. User (and application) behavior results in several packets disseminated over the network. An approach based on more general information, like network traffic, needs  to
map low-level information from network packets to higher-level user actions (for instance, brew a coffee or change the smart lock settings) in order to capture user behavior. This creates a challenge, as a single action could generate many packets addressed to different external destinations. \added{We need to design our system to be able to extract information about the user actions in this heterogeneous ecosystem to be able to model the behavior.}

\myparagraph{Capture User-Level Semantics} Our system is intended to map signals observed in the monitored data to higher-level semantic actions. This is challenging due to the semantic gap introduced by monitoring network traffic,   which is missing the context around user actions.

\myparagraph{Measurable and continuous authentication} Most systems for access control and authorization need to make an ultimate decision as to whether a user is allowed to access a critical resource or not. Our goal is to create an authentication service that generates \emph{authentication scores on a continuous basis}. These scores should provide a measure of authentication confidence, which is leveraged by higher-level authorization systems. For instance, to perform a high-value financial transaction, the authorization system could set the authentication score threshold at 0.95 and require another authentication factor (such as SMS). We believe that reliable authentication measures like the ones we aim to design can greatly augment the flexibility of current authorization systems.

%\end{enumerate}

\subsection{System Overview}
{\bf Our Approach.}  We address the challenges raised by our design goals by using the
IoT devices' network traffic that \added{passes through} the home router. In order to preserve user privacy, we collect the headers of the HTTPS packets that include minimal information such as timing, ports, destination address, \added{and} bytes sent and received, posing low risk on user privacy. Additionally, to protect user privacy we make the design choice of performing the model training and testing locally in the user's home, rather than remotely in the cloud. This implies that the authentication module needs to reside in the smart home. To address the mis-match between user-level semantics and network-level data, we \emph{observe the network traffic for a continuous time window} and design an ML model that uses features aggregated over a recent time window to predict the user in the room. Finally, we train a multi-class classifier to predict the likelihood that a certain user is in the room based on the most recent activities.

Figure \ref{fig:architecture} gives an overview of our system architecture. We installed 15 IoT devices in the IoT lab at our institution and  collect pcap files from them. We design an IRB-approved user study, with six users participating in data collection over a period of three weeks. During  training, we use labeled data of user sessions and extract features from HTTP headers over a continuous time window. We test multiple ML  algorithms with the aim of obtaining good accuracy at identifying users. During the regular user interaction with the IoT devices, the authentication module continuously receives data extracted from the most recent observation window and computes an authentication score for each user. The authentication score can be used by upper-level authorization systems for flexible policies according to various levels of risk. 
For example, \added{login} to a local device might require a lower score than performing a financial transaction in the cloud.

Below we provide more details about the machine learning models and the features we select for our  classifier.

\subsection{Machine Learning Design for User Classification}

We design a machine learning (ML) system that: (1) learns user profiles over time, using data extracted from network traffic generated by IoT devices; and (2) computes an authentication score in real time based on the most recent observation interval. The authentication score estimates the confidence or probability of the actual user being in the room. The ML system is trying to learn profiles for a set of users $\users=\{U_1,\dots, U_m\}$. In our specific application, user observation is performed over time, and thus the features or user attributes need to be defined over an observation window. We design the system to predict at time $t$ the probability that a certain user is in the room, i.e.,:
$$A^t_i = Pr[U_i^t|F(t,\Delta)], \forall i \in [1,m]$$

Given feature set $F(t,\Delta)$ computed over the time window $[t-\Delta,t]$ of size $\Delta$ (most recent history of length $\Delta$), the system aims to predict an authentication score $A_i^t$ at time $t$ for each user $U_i$. This can be achieved by training a multi-class classifier $f \in \Hcal$,  where $\Hcal$ is the hypothesis space. The ML algorithm is given historical training data of labeled user activities: $\dtrain=\{(F_1(t_1,\Delta),L_1),\dots,(F_n(t_n,\Delta),L_n)\}$, with labels $L_i \in \users$ and $F_i(t_i,\Delta)$ \added{being} the set of features computed over observation intervals $[t_i,t_i-\Delta]$ of length $\Delta$. A model $f$ is selected to minimize a loss function on the training set.
A probabilistic model in fact estimates the probability that a user generates the observed features in the most recent time window of length $\Delta$. The sum of the predicted probabilities is always 1:
\ignore{
$$\min_{f \in \Hcal} \ell(f(x_i^{\Delta},L_i,{\bf w})),$$
where ${\bf w}$ is the vector of model parameters learning during training.
}

$$\sum_{i=1}^m A_i^t = \sum_{i=1}^m Pr[U_i^t|F(t,\Delta)] = 1,\forall t $$
At testing time, the model is applied at time $t$ to the most recent observation window (i.e., features $F(t,\Delta)$ extracted during time interval $[t-\Delta,t]$) and generates an authentication score $A^t_i$ for each user $U_i \in \users$.
We experiment with multiple ML classification algorithms, including Logistic Regression, Random Forest, and Gradient Boosting that fit our framework very well. The most challenging issue in our design is to create appropriate feature representations that capture user behavior over the recent time window and can be used to effectively differentiate \added{between} multiple users with high confidence. %We discuss multiple feature representations next.

\subsection{Feature Representation}
\label{sec:features}

\myparagraph{Packet-level attributes} Our system design is based on features extracted from HTTPS traffic captured at the router in the smart home. To protect user privacy, one of our design considerations is to inspect only headers of the HTTPS packets. The fields that we leverage in our design include the device identifier (extracted from the MAC address in each packet), timestamp, the packet length, the direction (outgoing or incoming), transport-level protocol (TCP, UDP, or ICMP), destination port, source port, and destination domain.
%Since some of the HTTPS packets include directly IP addresses for external destinations%
We analyze the DNS queries from the network traffic, and extract mapping of external IP addresses to domain names (FQDN).

\myparagraph{Feature definition} Using packet-level attributes directly as features in an ML model is not feasible due to the large number of packets that are usually sent. Moreover, for our task of user authentication we need aggregated behavioral indicators, beyond the scope of a single network packet. Therefore, we need to experiment with multiple feature representations that capture the user behavior over a time interval  of length $\Delta$. For our approach, we use a sliding-window method to advance the time $t$ at which we compute authentication scores by one minute. We experiment with multiple values for the window length (from 5 to 30 minutes). For instance, if a user is in the room for an interval of 30 minutes during training and $\Delta=5$ minutes, we generate 26 sliding windows of size 5 minutes. Features are generated for each sliding window and labeled with the identity of the user. At testing time, we compute an authentication score at each time $t$ based on the \emph{most recent sliding window} of size $\Delta$, i.e., the interval $[t-\Delta,t]$. That means, implicitly, that we need to observe the user for \added{a period of} at least $\Delta$ before computing an authentication score.
The first source of data for feature extraction is device usage, usually a strong behavioral indicator. We define \emph{device-level features} extracted from packet attributes aggregated per device. In particular, we would like to capture information about user interaction with each IoT device during the time window $\Delta$, such as: (1) various statistics on incoming and outgoing packet lengths; (2) packet counts for transport-level protocols; (3) number of distinct domains contacted by each device; (4) statistics on inter-arrival timing of packets. 
%The list of aggregated features is given in Table~\ref{table_packet_stats}. 
All the features are aggregated over the packets sent and received by each device in each time window of size~$\Delta$. An important design decision is to separate incoming and outgoing traffic, as they might have different distributions. In particular, user interaction with voice assistants could result in large volumes of incoming traffic (for instance, if a user listens to music), while device communication with the cloud involves regular outgoing communication. We denote this set of device features in time window $[t-\Delta,t]$ as: $F(t,\Delta)^{\mathsf{dev}}$.
The second source of data useful for feature generation is user communication with external destinations (domain names). This is motivated by our observation that different Alexa skills connect to different external domains. For instance, listening to music may stream data from the {\sf spotify.com} domain, whereas playing fireplace sounds uses {\sf kwimer.com}. As the set of external domain names is very large and cloud services use many sub-domains for load balancing, we transform our domain names to second-level domains and compute features per second-level domain. For each second-level domain, we use aggregated features including statistics on incoming and outgoing packet lengths, inter-event times, and packet counts for transport protocols, computed over all the packets exchanged with that external domain in the time window of size $\Delta$. We denote this set of domain features in time window $[t-\Delta,t]$ as: $F(t,\Delta)^{\mathsf{dom}}$.

 It is well known that people have different interests and users prefer to interact with certain type of devices. Moreover, the same user tends to use the same set of IoT devices \added{that} they are most familiar with over time.  We believe that device usage is a strong behavioral indicator, as also confirmed in our evaluation.

We therefore consider several feature representations:

%\vspace{0.1pt}
%\begin{enumerate}
\noindent 1. \emph{Device-only features}: Predict $P[U_i|F(t,\Delta)^{\mathsf{dev}}]$;
\\\noindent 2. \emph{Domain-only features}: Predict $P[U_i|F(t,\Delta)^{\mathsf{dom}}]$;
\\\noindent 3. \emph{Device and domain features}: Predict $P[U_i|F(t,\Delta)^{\mathsf{dev}}; F(t,\Delta)^{\mathsf{dom}}]$.

% !TEX root = iot_auth_main.tex
\begin{table}[t]
	\small
	\caption{ IoT Devices installed in the IoT lab. During the user study, we investigate if users generate a unique signature over time with the choice of device, as well as different actions performed with the devices.}
	\begin{center}
			\scalebox{0.8}{
		\begin{tabular}{|c|c|}
			%{|p{0.30\textwidth}|p{0.50\textwidth}|}
			\hline
			{\bf Device} & {\bf User Activities} \\
			\hline
			Amazon Echo Dot, Spot, Plus & Alexa voice assistants \ignore{to support voice commands and control other smart devices} \\
			\hline
			
			Google Home Mini & Google voice assistant \ignore{to support voice commands and control other smart devices} \\
			\hline
			Harman Kardon Invoke & Cortana voice assistant \ignore{to support voice commands and control other smart devices} \\
			\hline
			
			LG Smart TV & Watch TV or stream content \ignore{through third-party applications} \\
			\hline
			
			Roku TV & Stream content \ignore{through third-party applications} \\
			\hline
			Amazon Fire TV & Stream content \ignore{through-third party applications} \\
			\hline
			Philips Hue Bridge & Set the color and brightness of the bulbs \\
			\hline
			
			Samsung Smart Fridge & Use fridge, interact with the LCD Display \\
			\hline
			Smart Microwave & Open, Heat up food, Close \\
			\hline
			
			iKettle Smart Kettle & Boil water\\
			\hline
			
			Behmor Smart Brewer & Brew coffee\\
			\hline
			SmartThings Hub & Trigger motion and contact sensors \\
			\hline
			
		\end{tabular}}
	\end{center}

	\label{table_activities}
	%	\vspace{-4mm}
\end{table}

\section{System Implementation}
\label{sec:data}

\subsection{User study}
\label{sec:userstudy}

We performed a user study with multiple users over three weeks to generate the dataset used to train and test our authentication models. We utilize the IoT lab at our institution to conduct our user study and monitor users while they interact with a smart-home like environment. The lab is an enclosed studio with a kitchen and living area, equipped with a range of Internet-connected devices and appliances to simulate a smart household. \ignore{The lab is available for use by researchers at our institution. }We collected data from 15 IoT devices installed in our IoT lab, as described in Table~\ref{table_activities}.

We asked users to use the room for multiple sessions, each of them lasting at least 15 minutes. One of our requirements was that users be alone in the room while we performed the user study to reduce interference from multiple users in training. Users logged their start and end times using their mobile phones in order to label the data for each session. We held several orientation sessions to help users become familiar with the devices. We did not provide any scripts, and users were asked to interact with the devices in a natural manner, similar to their actions at home.  
We recruited 10 users for the study, generating 74 sessions with a duration of 1660 minutes. 
%The usage statistics per user are in Table \ref{table_session_stats} in the Appendix.
We collected 4,082,975 packets, including connections to 316 external destinations and 97 second-level domains. In this paper, we focus on six users \added{(users 1, 3, 4, 6, 8, and 10),} who participated in more than 6 sessions.

\ignore{The reason is to facilitate data labeling and to ensure that we do not monitor data generated by other lab users. }

\myparagraph{IRB approval} Our study was approved by the IRB at our institution. All researchers with access to the collected data completed the IRB training, and users were required to consent to the data being collected while they used the room. Users were informed of the project goal before participating in the study. As we only extracted fields from packet headers of HTTPS traffic, we did not have access to any user personal information. To further protect the privacy of users, we anonymized the collected data by creating a unique user ID for each study participant. Our data is stored in anonymized format on our servers.

\myparagraph{Implementation}  The lab network is monitored at all times and the network traffic (pcap files) of all the IoT devices is collected in a server located in the room. We did not collect data from users' personal devices such as laptops and phones, as we used a MAC address filter to collect only IoT device traffic.  We created software that parses the HTTPS packet headers and stores the fields in a Postgres database. We extracted the device and domain features and stored them in a different table in our database.  For training our models, we used the ML implementation from the {\sf Python scikit-learn} package. We performed cross-validation for all our experiments and varied the ML hyper-parameters.

%to search for the best setting.

\begin{figure}[t]
	\begin{minipage}[b]{0.49\linewidth}
		\centering
		\includegraphics[width=\textwidth]{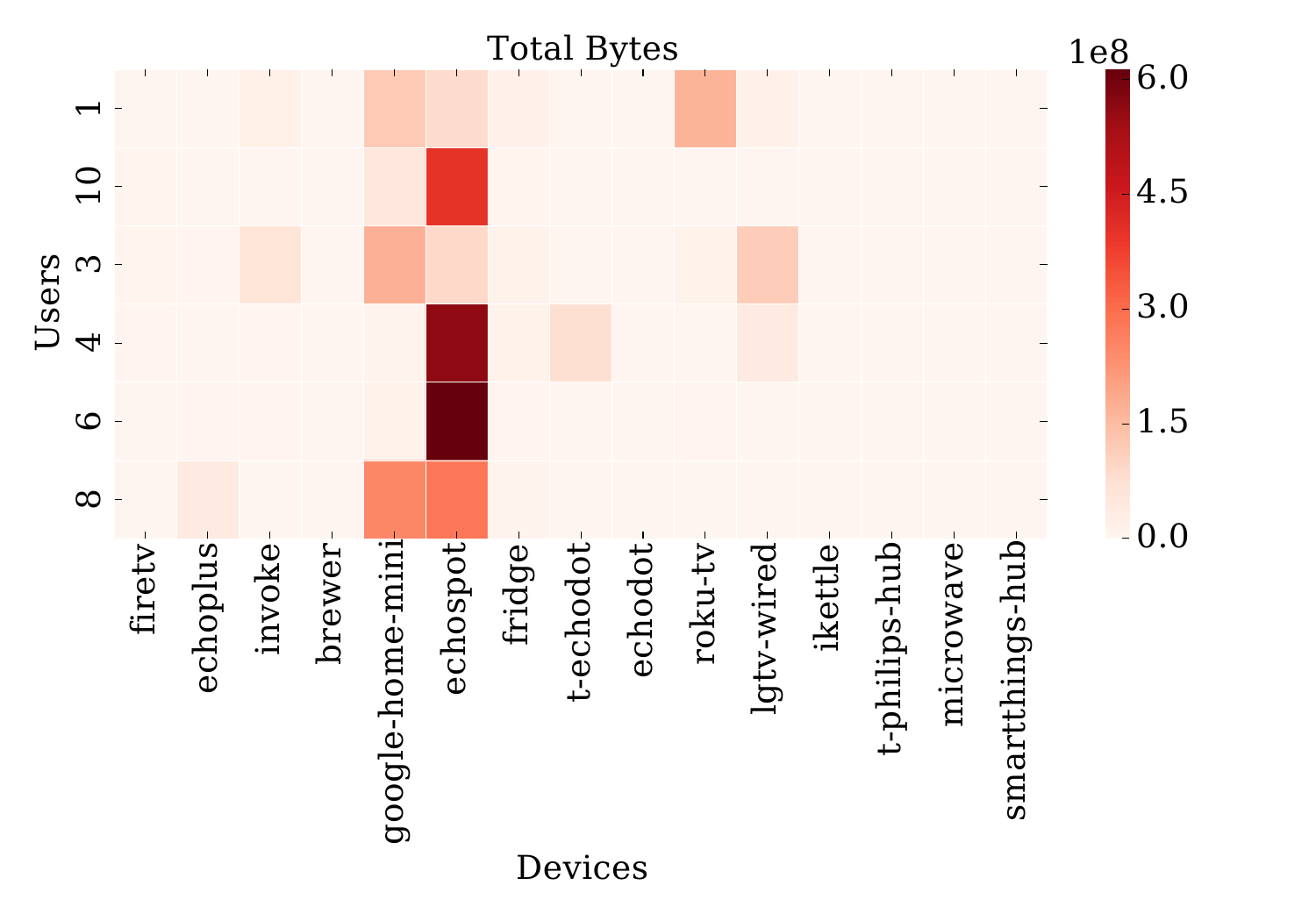}
		
	\end{minipage}
	%\hspace{0.5cm}
	\begin{minipage}[b]{0.49\linewidth}
		\centering
		\includegraphics[width=\textwidth]{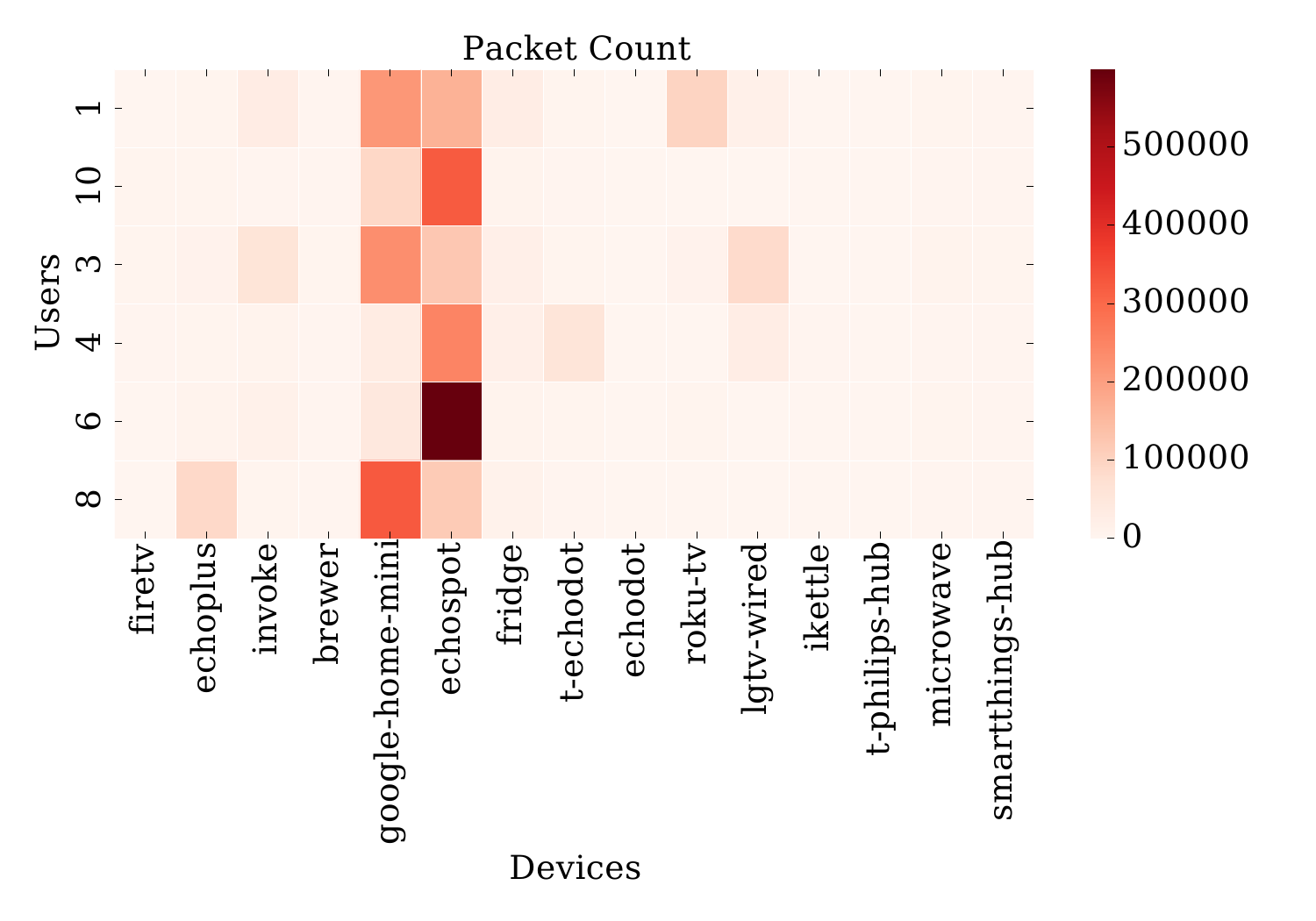}
		
	\end{minipage}

	\begin{minipage}[b]{0.49\linewidth}
		\centering
		\includegraphics[width=\textwidth]{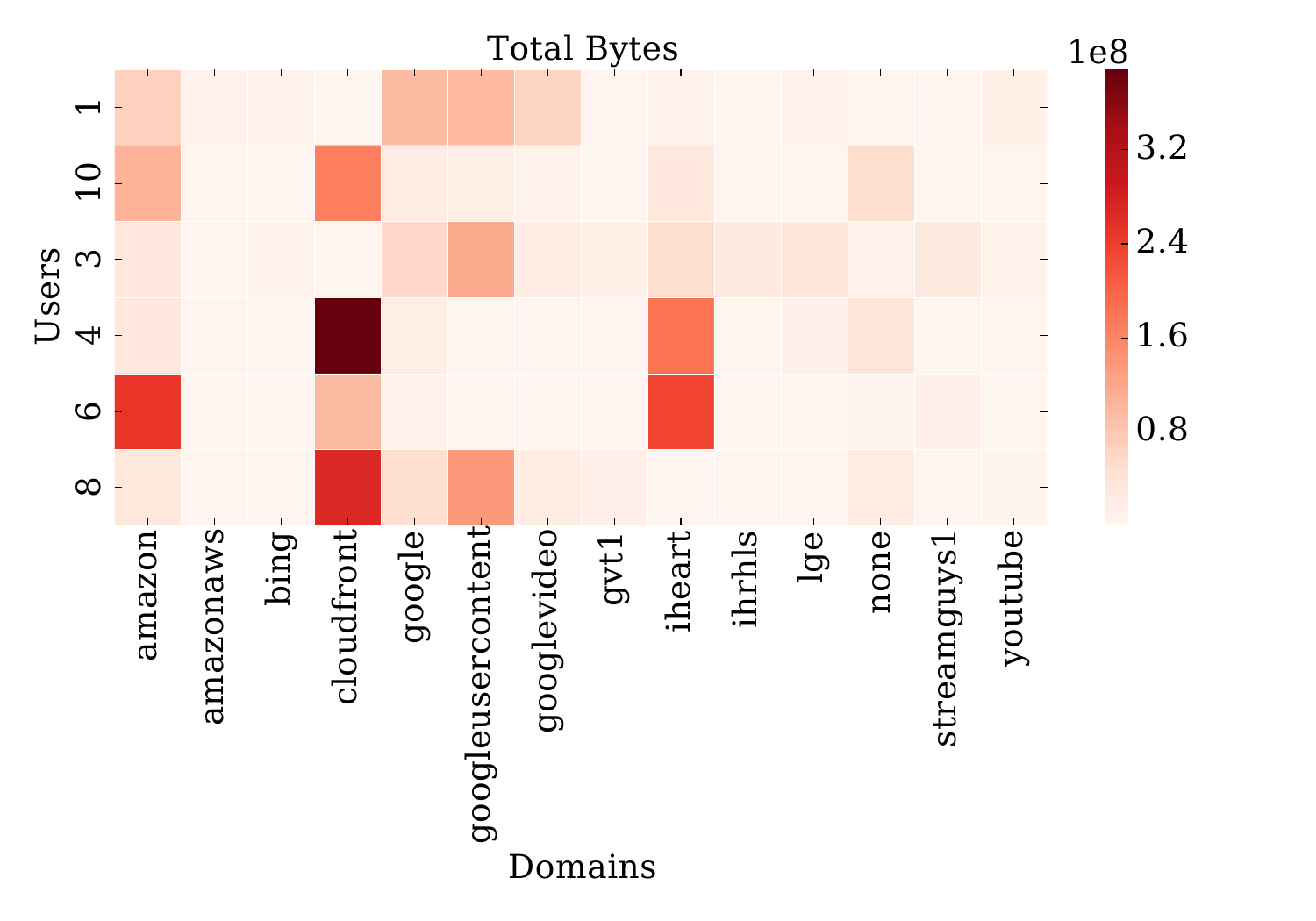}
		
	\end{minipage}
	%\hspace{0.5cm}
	\begin{minipage}[b]{0.49\linewidth}
		\centering
		\includegraphics[width=\textwidth]{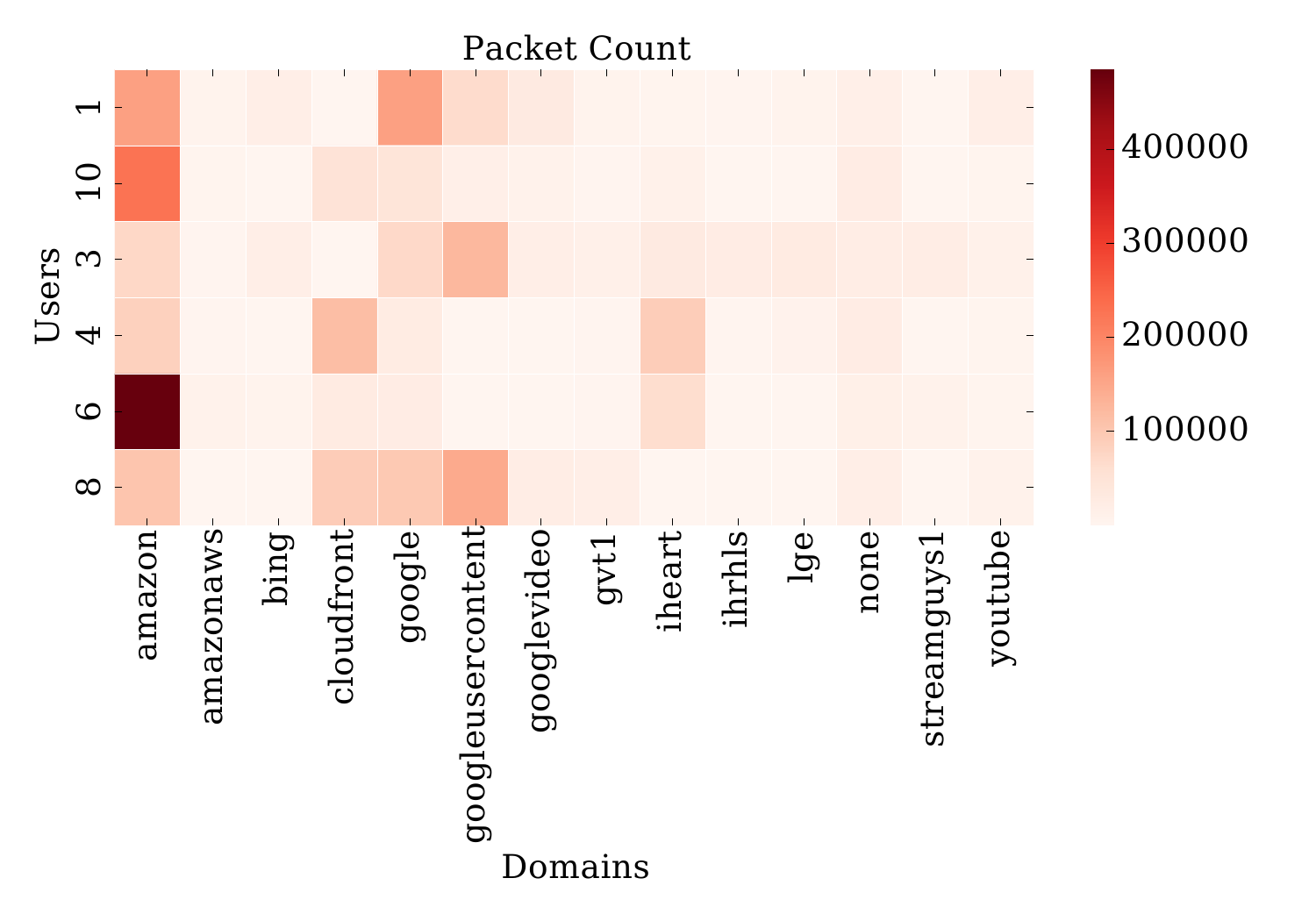}
		
	\end{minipage}
	%\hspace{0.5cm}
	%		\vspace{-4mm}
	%\captionsetup{justification=centering,margin=0.5cm}
	\caption{Device (top) and Domain (bottom) usage statistics during the user study. Total bytes (left), and packet count (right) for each user demonstrate common and unique identifiers in terms of devices and domains that correspond to different actions.}
	\label{fig:figure_domainusage}
				\vspace{-4mm}
\end{figure}

\begin{figure*}[t]
	\centering
	\begin{minipage}[b]{0.3\linewidth}
		\centering
		\includegraphics[width=\textwidth]{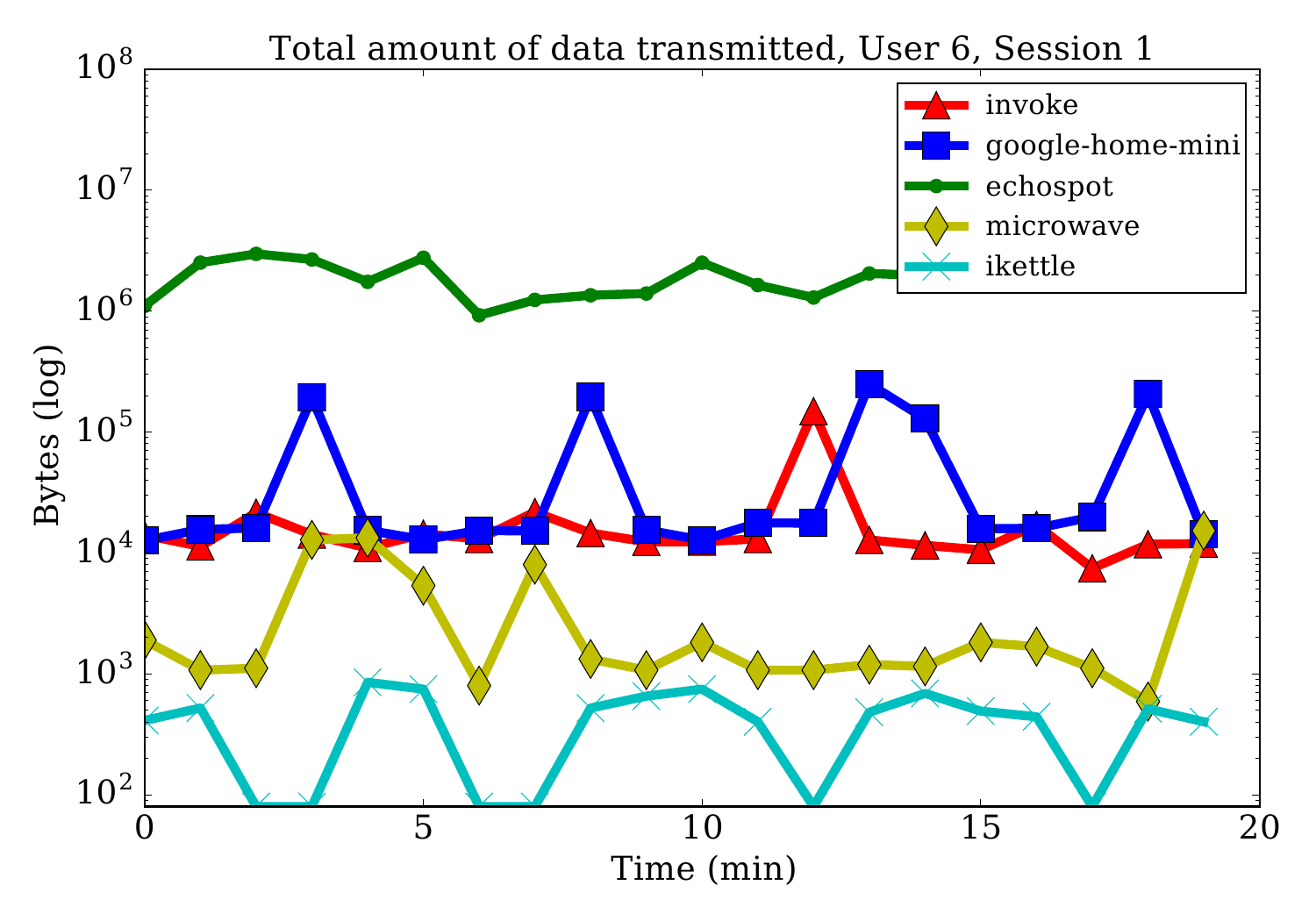}
		
	\end{minipage}
	%\hspace{0.5cm}
	\begin{minipage}[b]{0.3\linewidth}
		\centering
		\includegraphics[width=\textwidth]{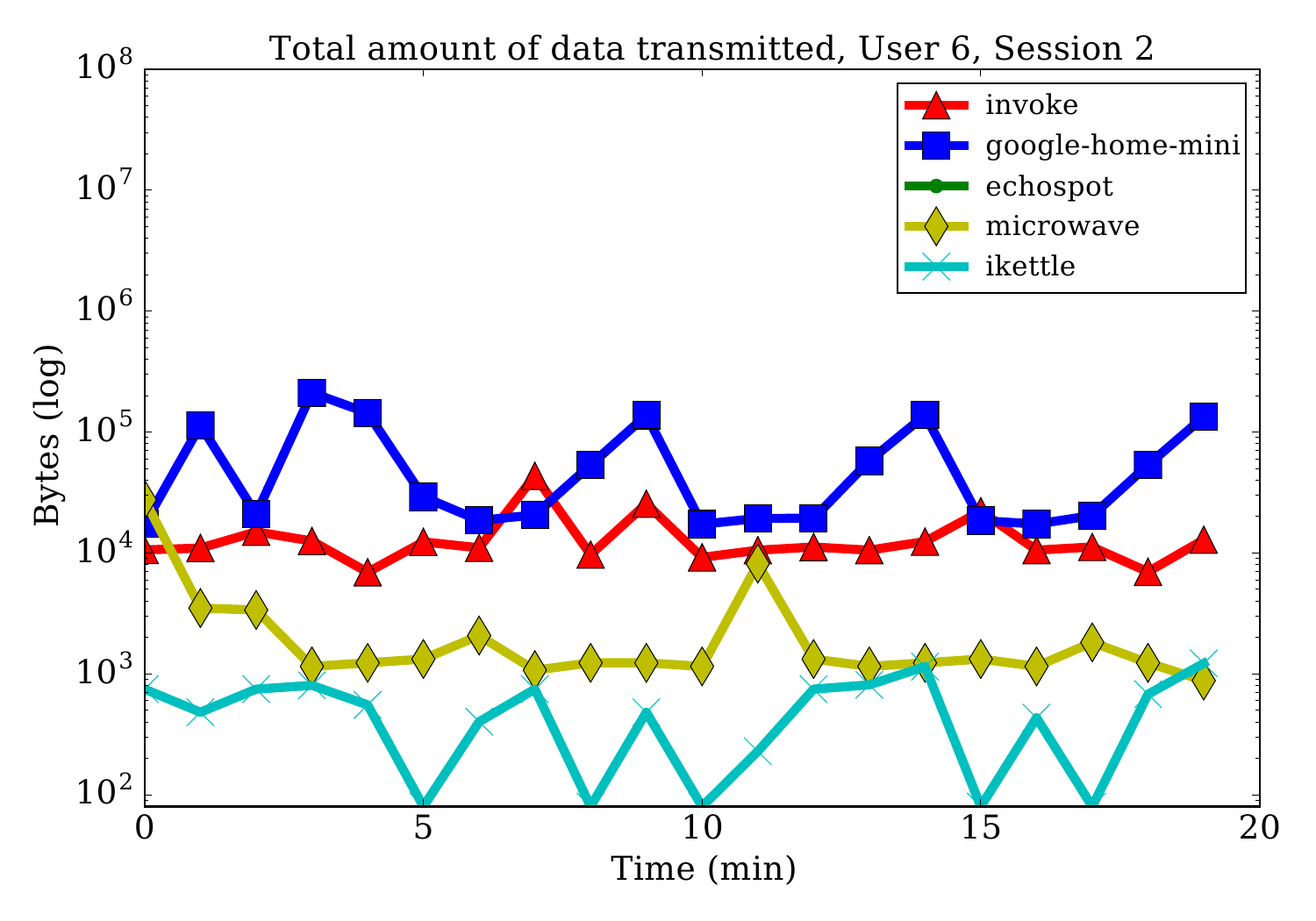}
		
	\end{minipage}
	%\hspace{0.5cm}
	\begin{minipage}[b]{0.3\linewidth}
		\centering
		\includegraphics[width=\textwidth]{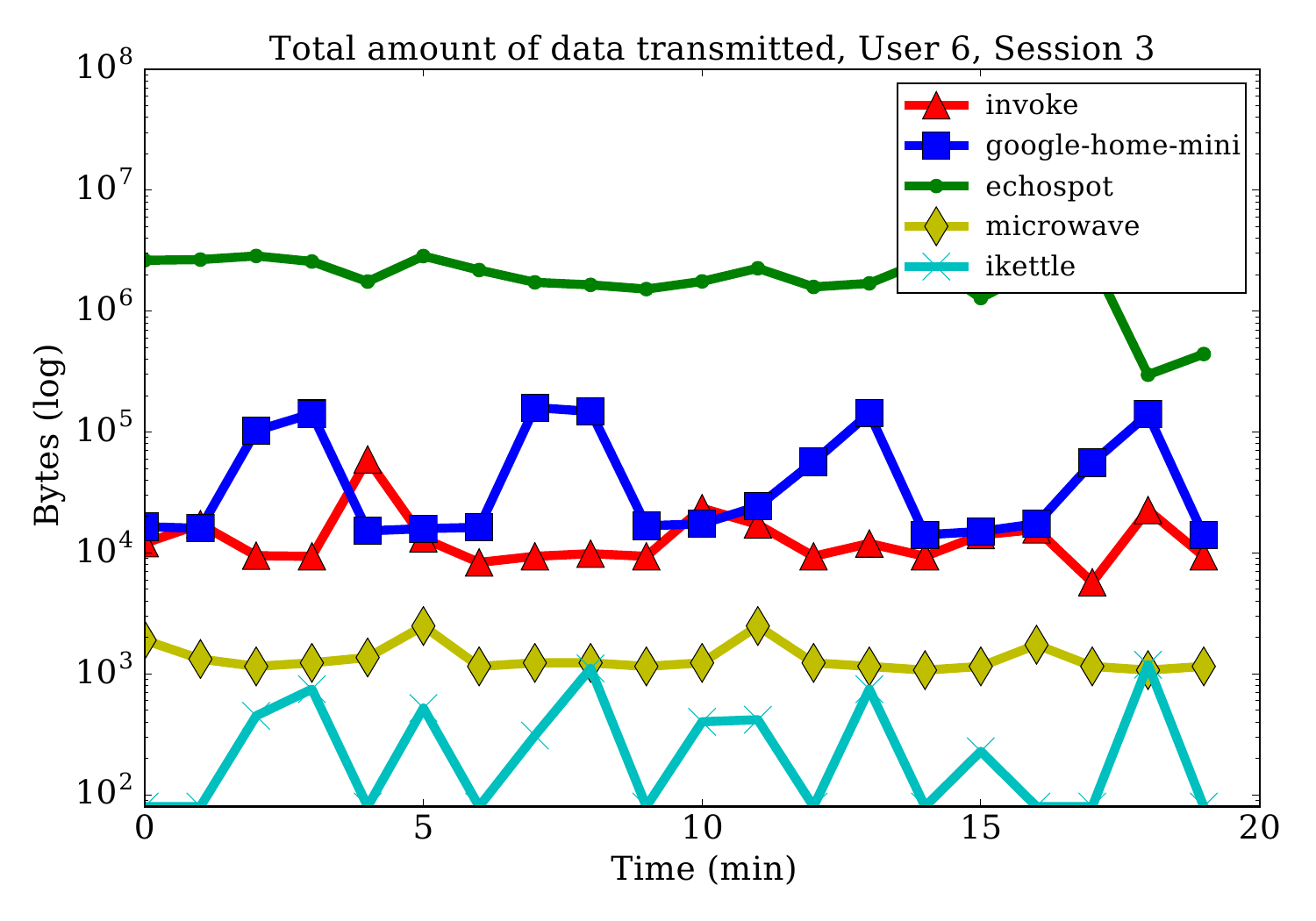}
		
	\end{minipage}
		%	\vspace{-4mm}
%	\captionsetup{justification=centering,margin=0cm}
%	\caption{Data transmitted per device for User 6 in three sessions.}
%	\label{fig:figure_user6_sessions}

	\centering
	\begin{minipage}[b]{0.3\linewidth}
		\centering
		\includegraphics[width=\textwidth]{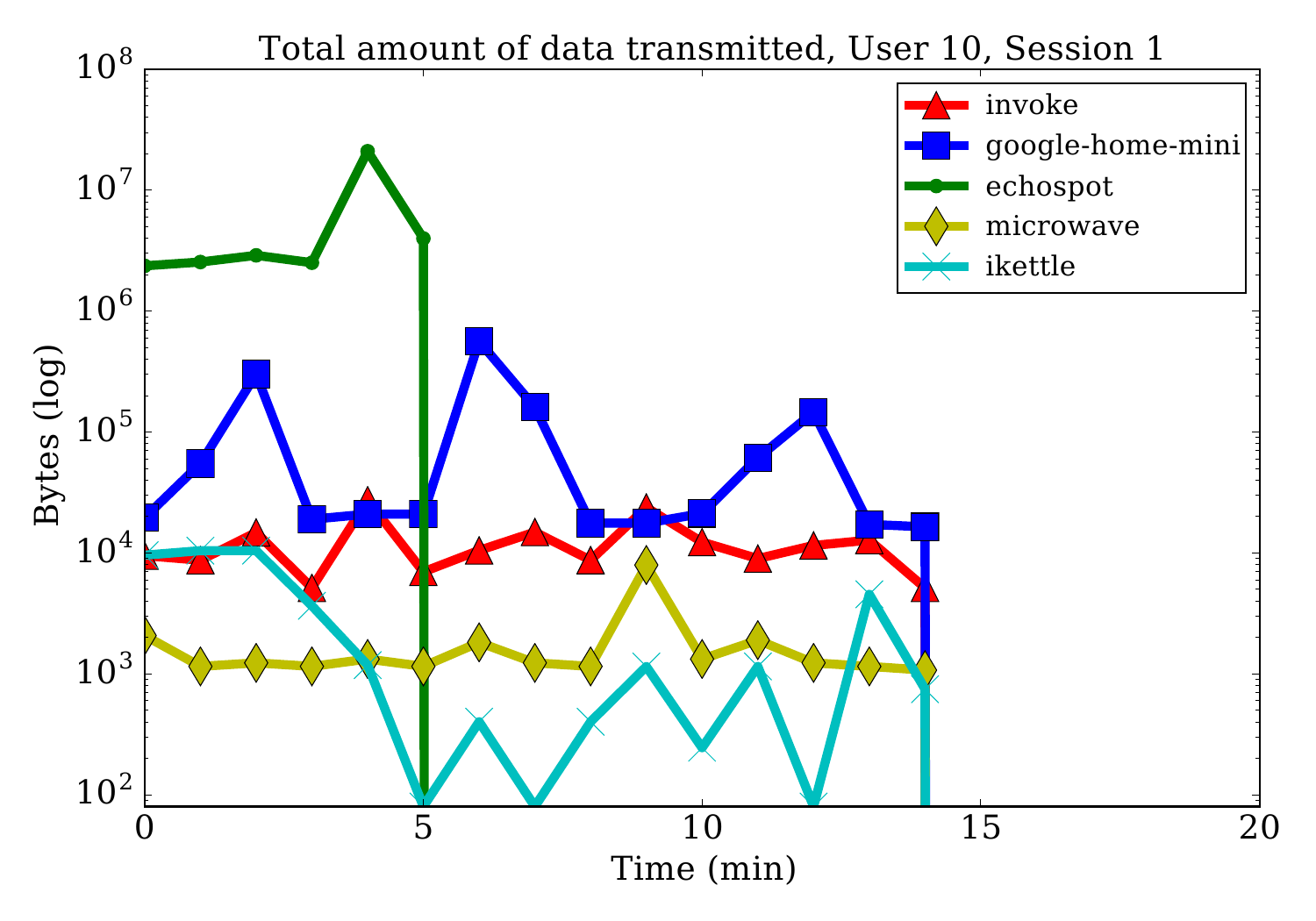}
		
	\end{minipage}
%	\hspace{0.5cm}
	\begin{minipage}[b]{0.3\linewidth}
		\centering
		\includegraphics[width=\textwidth]{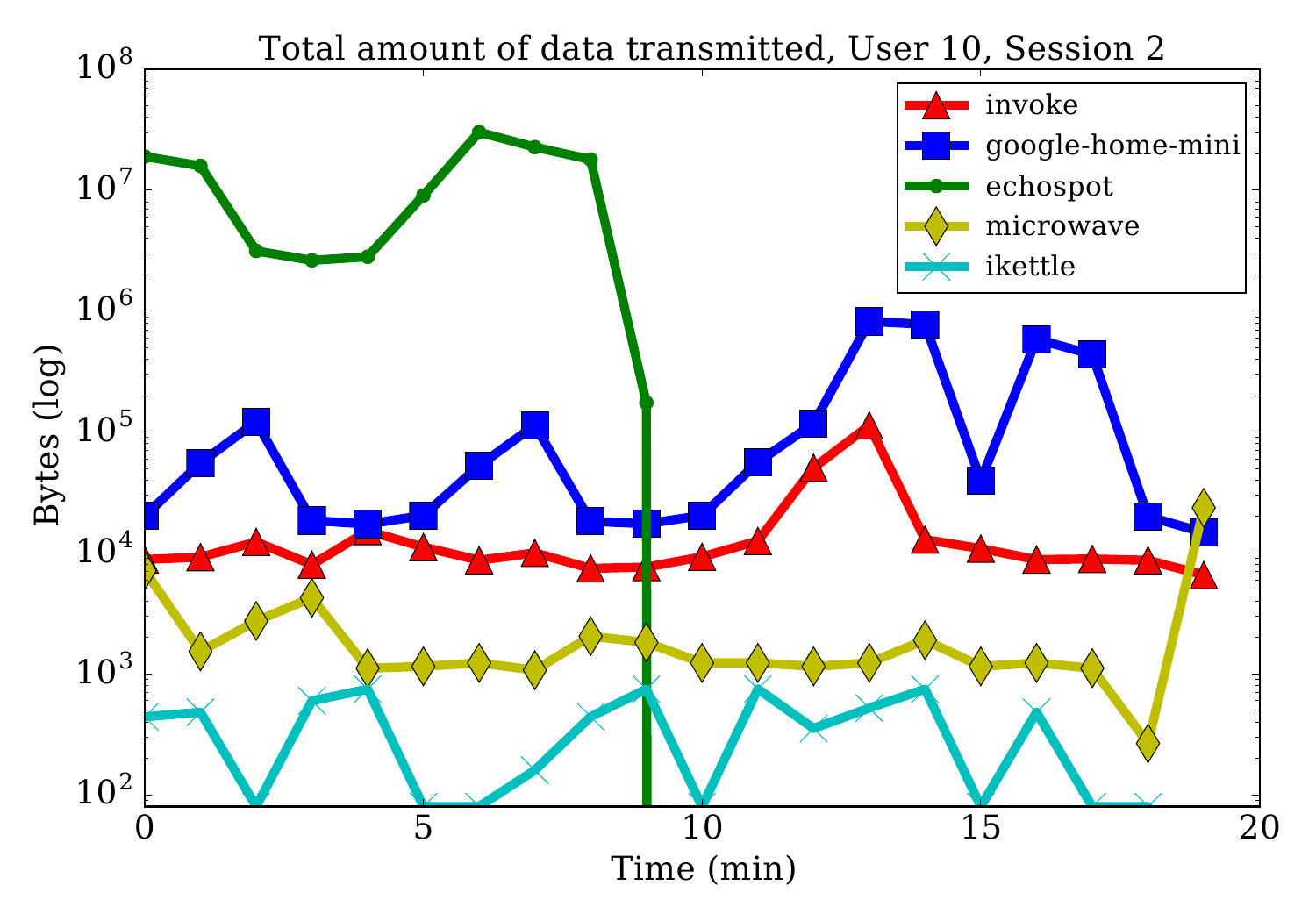}
		
	\end{minipage}
%	\hspace{0.5cm}
	\begin{minipage}[b]{0.3\linewidth}
		\centering
		\includegraphics[width=\textwidth]{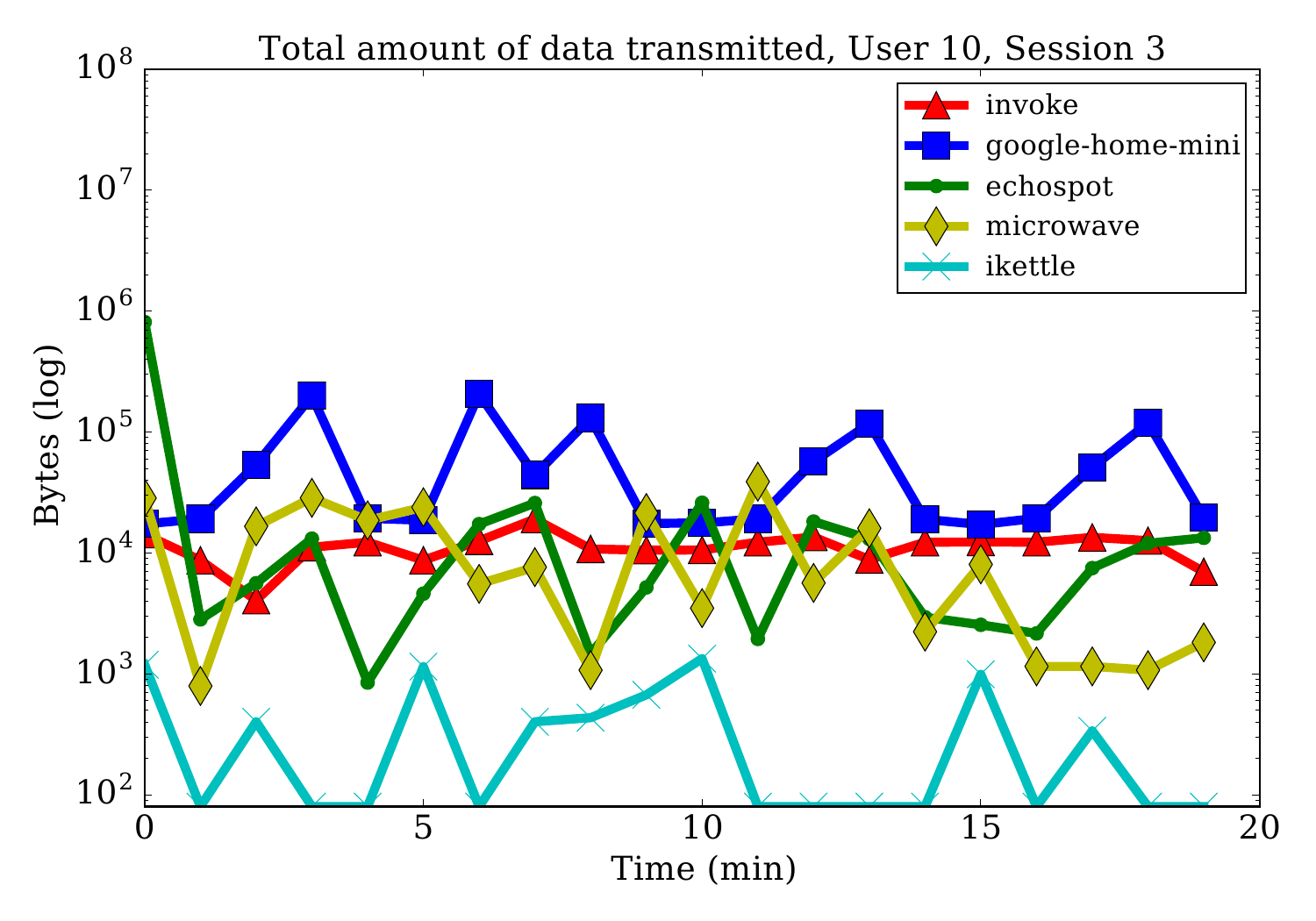}
		
	\end{minipage}
			%\vspace{-4mm}
	\captionsetup{justification=centering,margin=0cm}
	\caption{Data transmitted per device for User 6 (top) and 10 (bottom) in three sessions. While there is some similarity across the same user's sessions, these plots show user behavior varying across sessions, making behavioral authentication in this context challenging.}
	\label{fig:figure_user10_sessions}
	
\end{figure*}

\subsection{Data exploration} We analyzed the collected data  to compare different users in terms of device and domain usage. In top row of Figure \ref{fig:figure_domainusage}, we show the total data exchanged, the number of packets by device and user. For some devices such as Amazon Echo Spot and Google Home, the amount of data transferred varies significantly by user. On the other hand,  some smart devices (such as the microwave and smart kettle) are either not utilized by users, or they do not generate a lot of network activity. 

In bottom row of Figure \ref{fig:figure_domainusage}, we display the amount of data transferred, the total number of packets by user and second-level domains. We only included domains with more than 10,000 packets for at least one user \added{for visualization purposes, and in the rest of the experiments we included all the domains.} While multiple devices could communicate with the same cloud and CDN domains (e.g., {\sf cloudfront.com}), some domains (such as {\sf iheart.com}) map to a more specific behavior, and exhibit different patterns across users.
We are interested in validating the hypotheses that users behave consistently  and users have differentiating behavior from other users. For this, we plot the amount of data transmitted by different devices over time for User 6 and User 10 in Figure~\ref{fig:figure_user10_sessions} in three distinct sessions. Interestingly, some sessions of the same user are  similar (for instance, the first and third session of User 6). However, users deviate in their behavior across different sessions, as demonstrated by the third session of User 10 (in which interaction with the Amazon Echo Dot is lower). Overall, Users 6 and 10 have different behavior in their interactions with  devices. While there is some similarity across the same user's sessions, these plots show user behavior varying across sessions, making behavioral authentication in this context challenging.

% !TEX root = iot_auth_main.tex

\begin{figure*}[t]
	\centering
	\begin{minipage}[b]{0.29\linewidth}
		\centering
		\includegraphics[width=\textwidth]{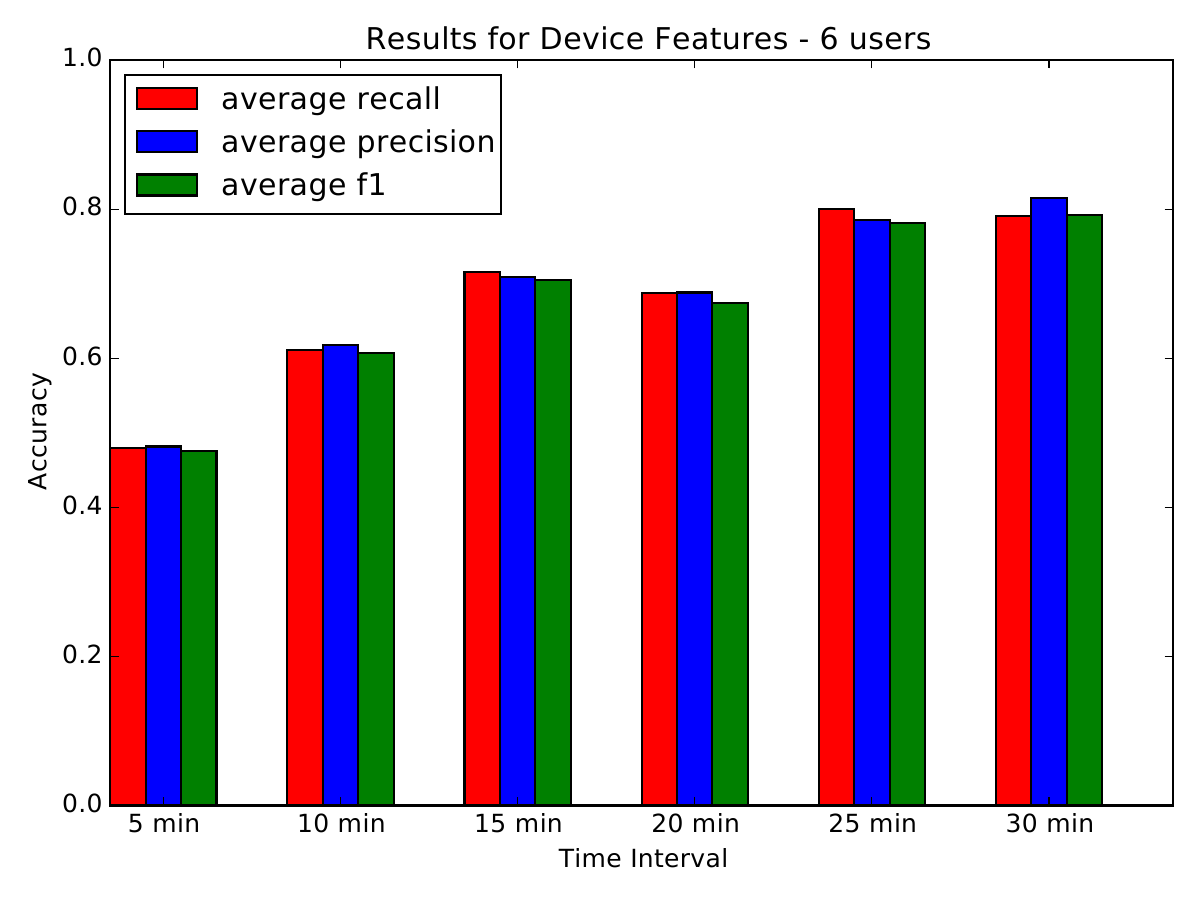}
		
	\end{minipage}
	%\hspace{0.5cm}
	\begin{minipage}[b]{0.29\linewidth}
		\centering
		\includegraphics[width=\textwidth]{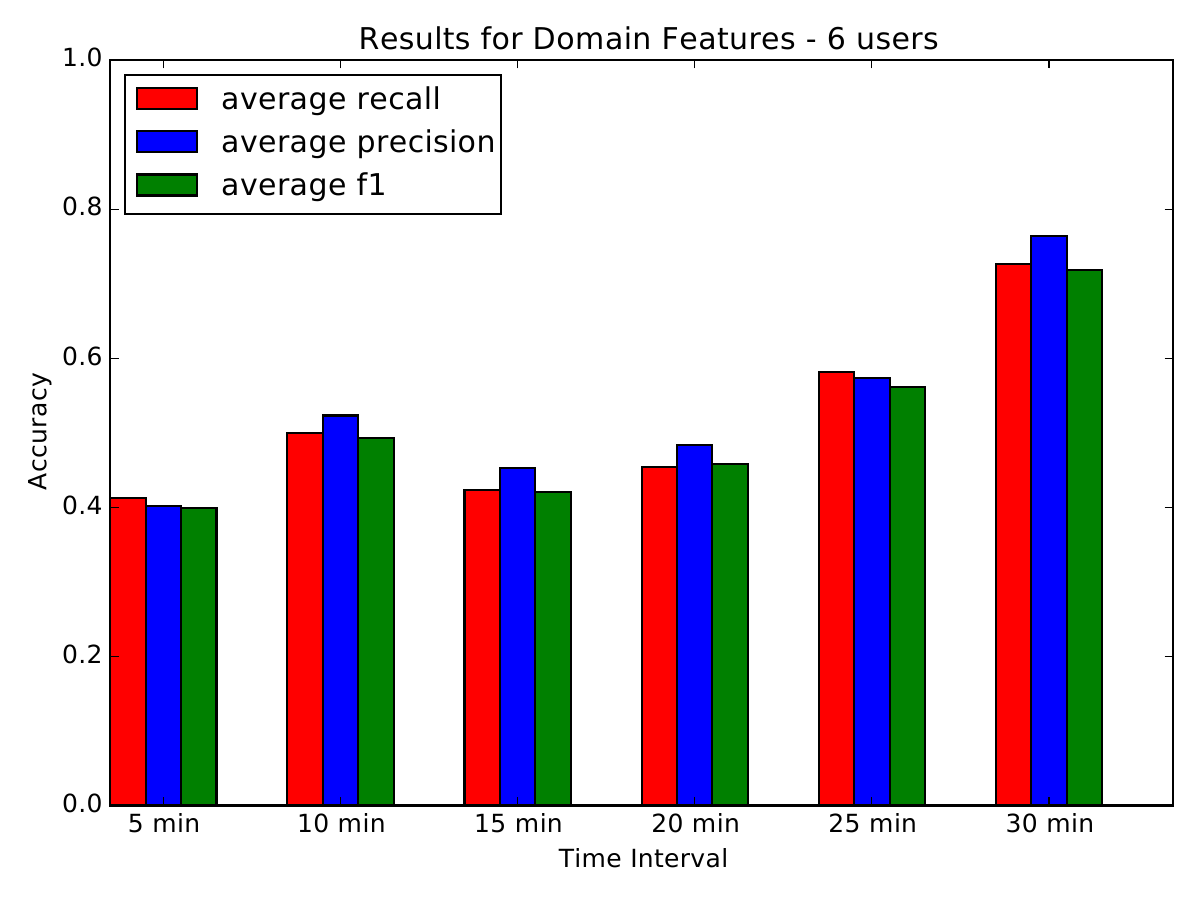}
		
	\end{minipage}
	\begin{minipage}[b]{0.29\linewidth}
		\centering
		\includegraphics[width=\textwidth]{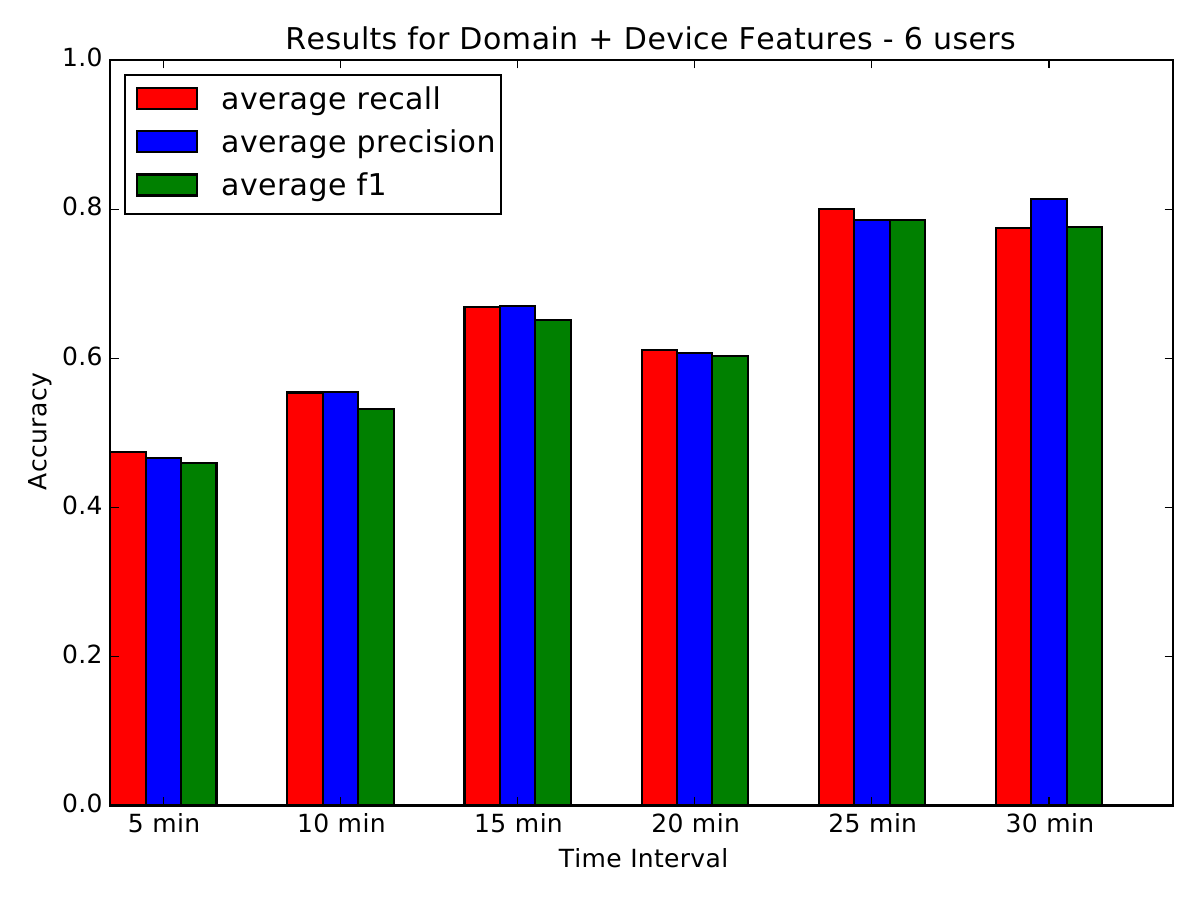}
		
	\end{minipage}
%			\vspace{-4mm}
	%\captionsetup{justification=centering,margin=0cm}
	\caption{Accuracy metrics for different time windows using three feature representations: device-only, domain-only, and combined features. Performance  generally increases as the time window increases, and device features are stronger than domain features.}
	\label{fig:figure_accuracy_6_users}
		
\end{figure*}

\section{Evaluation}
\label{sec:evaluation}

We evaluate our system using the data collected in the user study that is discussed in Section~\ref{sec:userstudy}. We present results for different feature representations, user classification performance, and receiver operating characteristic (ROC) curves for three different classification models and a high-confidence ensemble model.

\subsection{Comparison of Feature Representations}

We experiment with three different feature representations as discussed in Section~\ref{sec:features}: \emph{Device-only} (420 features); \emph{Domain-only} (2910 features); and \emph{Both device and domain} (3330 features).  We considered time windows of different lengths $\Delta \in \{5, 10, 15, 20, 25, 30\}$ minutes. Each time window is created by sliding windows over the user sessions at one-minute intervals. If a session is shorter than $\Delta$, we generate only one window covering the entire session.
As it is usually not possible to train accurate machine learning models with a limited amount of data, we decided to exclude users with fewer than 6 sessions in total. We perform our analysis on six users: 1, 3, 4, 6, 8, and 10 who participated in more than 6 sessions.
To evaluate our ML models, we perform 7-fold cross-validation across sessions. Thus, we keep one session for each user for testing and select all other sessions in training. We consider first a Random Forest (RF) classifier to evaluate feature representations and time window length. RF is an ensemble learning method that constructs many decision trees using a subset of features chosen at random at each split. Random Forests are robust models that work well in multi-class classification settings with many features. We will experiment and compare with other models in the next section.
%As the minimum number of sessions per user is 7, we perform 7-fold cross validation.
%Moreover, it enables us to investigate the feature importance to be able to explain the decisions made by the model and map them with user behavior.
We plot the recall, precision, and F1 score of the three feature representation methods for varying time windows in Figure~\ref{fig:figure_accuracy_6_users}. The results show that the model performance  generally increases as the time window increases until 25 minutes, at which point the best results are obtained. We thus select a time window of 25 minutes for the rest of our experiments. We note the inherent tradeoffs between  classification accuracy and the length of observation period for computing authentication scores. With smaller windows length, we can generate authentication scores after limited observation, at the cost of decreased classification performance. Comparing different feature representations, we observe that \emph{Device-only} performs well, and it is very similar to \emph{Both device and domain}, and device features are stronger than domain features.
We also generate the user confusion matrix for the RF models in Table~\ref{table:confusion_6user_25min}.
Looking at the confusion matrix for RF model, we obtain strong classification for some of the users (e.g., 0.93 F1 score for User 4), but worse results for other users (e.g., 0.51 F1 score for User 6). User 6 is mis-classified as User 3, meaning that their behavior is fairly similar. Thus, the classification performance varies \added{significantly} across users. We believe there are multiple factors contributing to this phenomenon, among them the variability of user behavior and the amount of data used for training these models.

\begin{table}[thb]
	\centering
	\small
		\caption{RF user confusion matrix and accuracy metrics for 25 minute intervals: avg. recall=$0.8$, avg. precision=$0.78$. We obtain very good classification for some of the users (e.g., 0.93 F1 score for User 4), but worse results for other users (e.g., 0.51 F1 score for User 6). User 6 is mis-classified as User 3, meaning that their behavior is fairly similar.}
	
	\scalebox{0.85}{
			\renewcommand{\arraystretch}{1.2}
	\begin{tabular}{|c|c|c|c|c|c|c|c|c|c|c|}
		\hline
		& \textbf{U1} & \textbf{U3} & \textbf{U4} & \textbf{U6} & \textbf{U8} & \textbf{U10} & \textbf{Count} & \textbf{Rec} & \textbf{Prec} & \textbf{F1} \\ \hline
		\textbf{U1}  & 5          & 0          & 1          & 1          & 0          & 0           & 7              & 0.71            & 0.71               & 0.71        \\ \hline
		\textbf{U3}  & 0          & 47         & 0          & 1          & 0          & 0           & 48             & 0.97            & 0.81               & 0.88        \\ \hline
		\textbf{U4}  & 0          & 1          & 20         & 1          & 0          & 0           & 22             & 0.90            & 0.95               & 0.93        \\ \hline
		\textbf{U6}  & 0          & 9          & 0          & 4          & 0          & 1           & 14             & 0.28            & 0.57               & 0.38        \\ \hline
		\textbf{U8}  & 1          & 0          & 0          & 0          & 4          & 2           & 7              & 0.57            & 0.8                & 0.66        \\ \hline
		\textbf{U10} & 1          & 1          & 0          & 0          & 1          & 4           & 7              & 0.57            & 0.57               & 0.57        \\ \hline
	\end{tabular}}
	\label{table:confusion_6user_25min}
\end{table}

\begin{table}[!t]
	\centering
	\small
			\caption{Comparison of three ML classifiers  for 6 users (top) and 5 users (bottom). GradientBoosting outperforms other models in both settings. }
	\scalebox{0.87}{
		\renewcommand{\arraystretch}{1.2}
		\begin{tabular}{|c|c|c|c|c|c|c|c|c|c|ll}
			\cline{1-10}
			\multirow{2}{*}{} & \multicolumn{3}{c|}{\textbf{LogisticRegression}}    & \multicolumn{3}{c|}{\textbf{RandomForest}}          & \multicolumn{3}{c|}{\textbf{GradientBoost}}      &  &  \\ \cline{2-10}
			& \textit{Rec} & \textit{Prec} & \textit{F1}   & \textit{Rec} & \textit{Prec} & \textit{F1}   & \textit{Rec} & \textit{Prec} & \textit{F1}   &  &  \\ \cline{1-10}
			\textbf{U 1}   & 0.42            & 0.6                & 0.5           & 0.71            & 0.71               & 0.71          & \textbf{0.85}   & \textbf{0.75}      & \textbf{0.79} &  &  \\ \cline{1-10}
			\textbf{U 3}   & 0.52            & 0.69               & 0.59          & \textbf{0.97}   & 0.81               & 0.88          & 0.93            & \textbf{0.86}      & \textbf{0.9}  &  &  \\ \cline{1-10}
			\textbf{U 4}   & 0.81            & 0.62               & 0.70          & 0.90            & 0.95               & 0.93          & \textbf{0.90}   & \textbf{1.0}       & \textbf{0.95} &  &  \\ \cline{1-10}
			\textbf{U 6}   & 0.71            & 0.45               & 0.55          & 0.28            & \textbf{0.57}      & 0.38          & \textbf{0.5}    & 0.53               & \textbf{0.51} &  &  \\ \cline{1-10}
			\textbf{U 8}   & 0.42            & 0.5                & 0.46          & \textbf{0.57}   & \textbf{0.8}       & \textbf{0.66} & 0.42            & 0.75               & 0.54          &  &  \\ \cline{1-10}
			\textbf{U 10}  & \textbf{0.71}   & \textbf{0.71}      & \textbf{0.71} & 0.57            & 0.57               & 0.57          & 0.57            & 0.5                & 0.53          &  &  \\ \cline{1-10}
			\textbf{Avg}  & 0.60            & 0.62               & 0.60          & 0.80            & 0.78               & 0.78          & \textbf{0.80}   & \textbf{0.81}      & \textbf{0.80} &  &  \\ \cline{1-10}
	\end{tabular}}

	\centering
	\small

	\scalebox{0.87}{
		\renewcommand{\arraystretch}{1.2}
		\begin{tabular}{|c|c|c|c|c|c|c|c|c|c|ll}
			\cline{1-10}
			\multirow{2}{*}{} & \multicolumn{3}{c|}{\textbf{LogisticRegression}}    & \multicolumn{3}{c|}{\textbf{RandomForest}}        & \multicolumn{3}{c|}{\textbf{GradientBoost}}      &  &  \\ \cline{2-10}
			& \textit{Rec} & \textit{Prec} & \textit{F1}   & \textit{Rec} & \textit{Prec} & \textit{F1} & \textit{Rec} & \textit{Prec} & \textit{F1}   &  &  \\ \cline{1-10}
			\textbf{U 1}   & 0.57            & 0.66               & 0.61          & 0.85            & 0.75               & 0.8         & \textbf{0.85}   & \textbf{1.0}       & \textbf{0.92} &  &  \\ \cline{1-10}
			\textbf{U 3}   & 0.79            & 0.88               & 0.83          & 0.97            & 0.95               & 0.96        & \textbf{1.0}    & \textbf{0.92}      & \textbf{0.96} &  &  \\ \cline{1-10}
			\textbf{U 4}   & 0.90            & 0.62               & 0.74          & 0.90            & 0.95               & 0.93        & \textbf{0.95}   & \textbf{1.0}       & \textbf{0.97} &  &  \\ \cline{1-10}
			\textbf{U 8}   & 0.42            & 0.75               & 0.54          & 0.57            & 0.8                & 0.66        & \textbf{0.71}   & \textbf{0.83}      & \textbf{0.76} &  &  \\ \cline{1-10}
			\textbf{U 10}  & \textbf{0.71}   & \textbf{0.83}      & \textbf{0.76} & 0.71            & 0.62               & 0.66        & 0.57            & 0.66               & 0.61          &  &  \\ \cline{1-10}
			\textbf{Avg}  & 0.76            & 0.79               & 0.78          & 0.90            & 0.90               & 0.90        & \textbf{0.92}   & \textbf{0.92}      & \textbf{0.92} &  &  \\ \cline{1-10}
	\end{tabular}}
	\vspace{1mm}
	\captionsetup{justification=centering,margin=0cm}

	\label{fig:figure_classifier_comparison}
	
\end{table}

\subsection{Model Comparison}

\begin{figure*}[t]
	\centering
	\begin{minipage}[b]{0.3\linewidth}
		\centering
		\includegraphics[width=\textwidth]{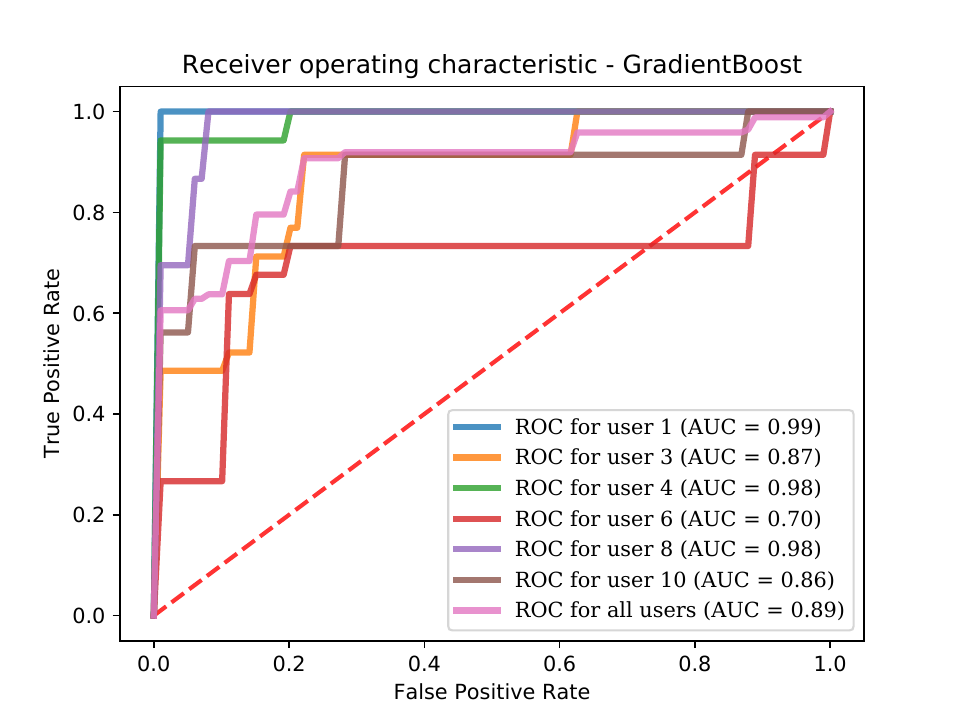}
		
	\end{minipage}
%	\hspace{0.1cm}
	\begin{minipage}[b]{0.3\linewidth}
		\centering
		\includegraphics[width=\textwidth]{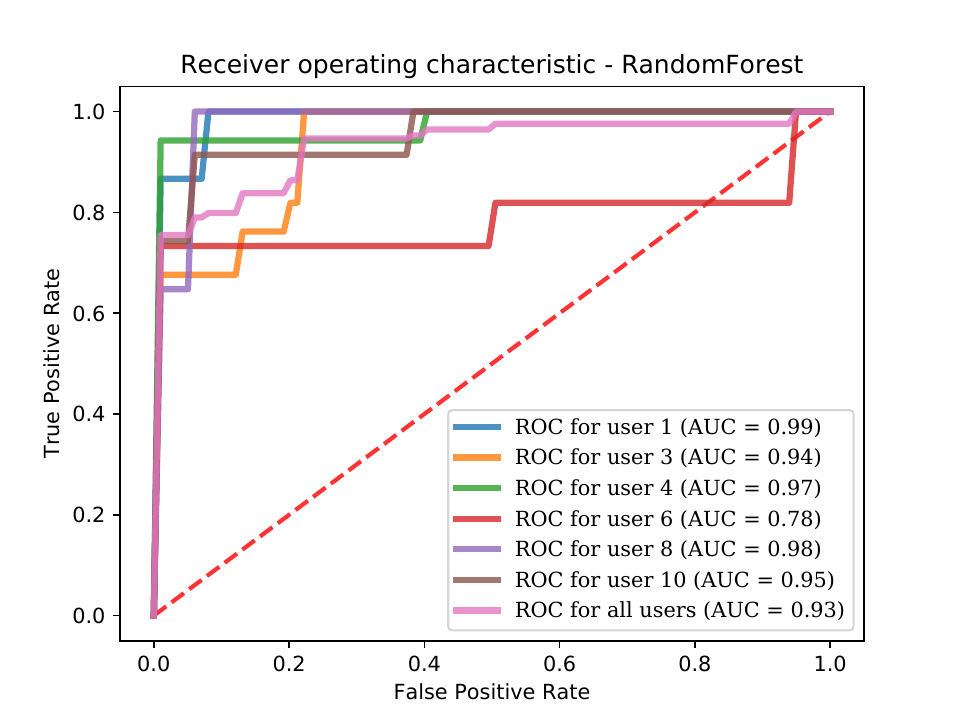}
		
	\end{minipage}
%	\hspace{0.1cm}
	\begin{minipage}[b]{0.3\linewidth}
		\centering
		\includegraphics[width=\textwidth]{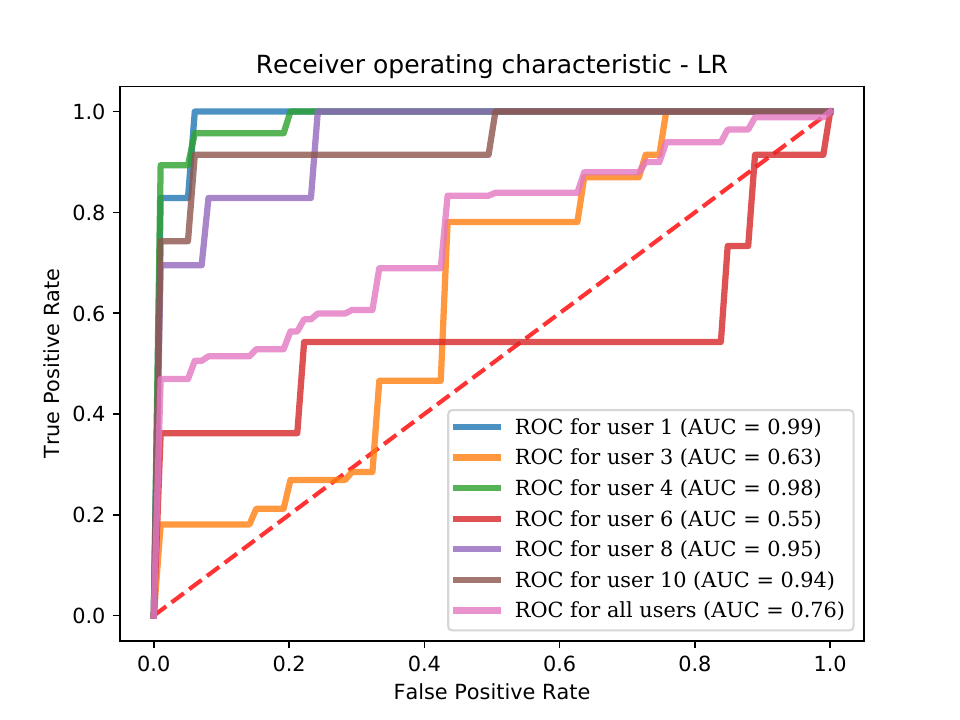}
		
	\end{minipage}

	\begin{minipage}[b]{0.3\linewidth}
		\centering
		\includegraphics[width=\textwidth]{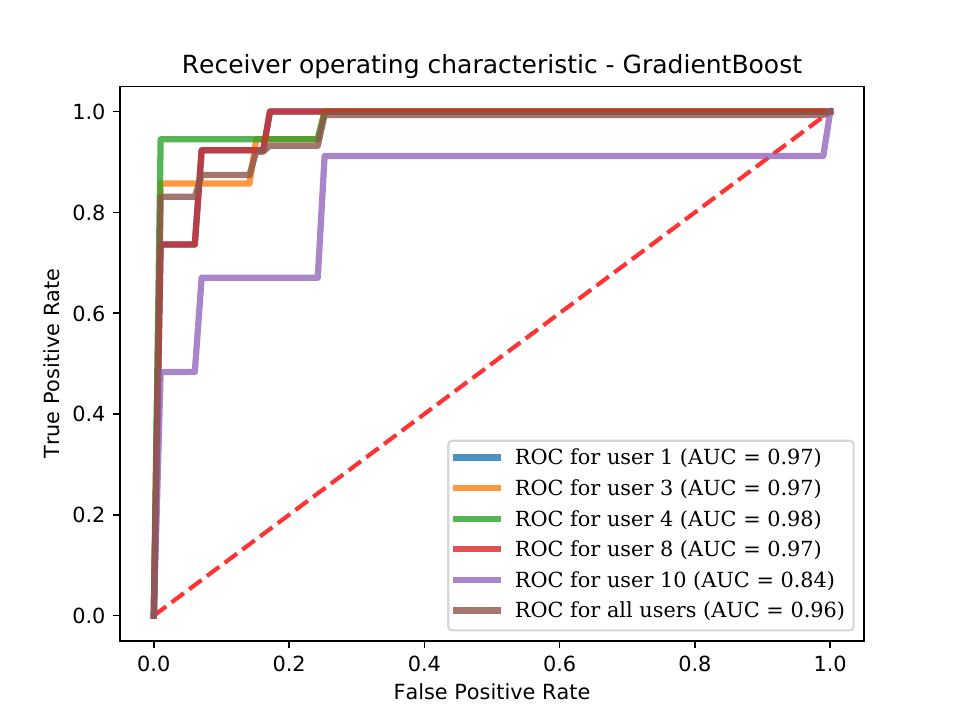}
		
	\end{minipage}
	%	\hspace{0.1cm}
	\begin{minipage}[b]{0.3\linewidth}
		\centering
		\includegraphics[width=\textwidth]{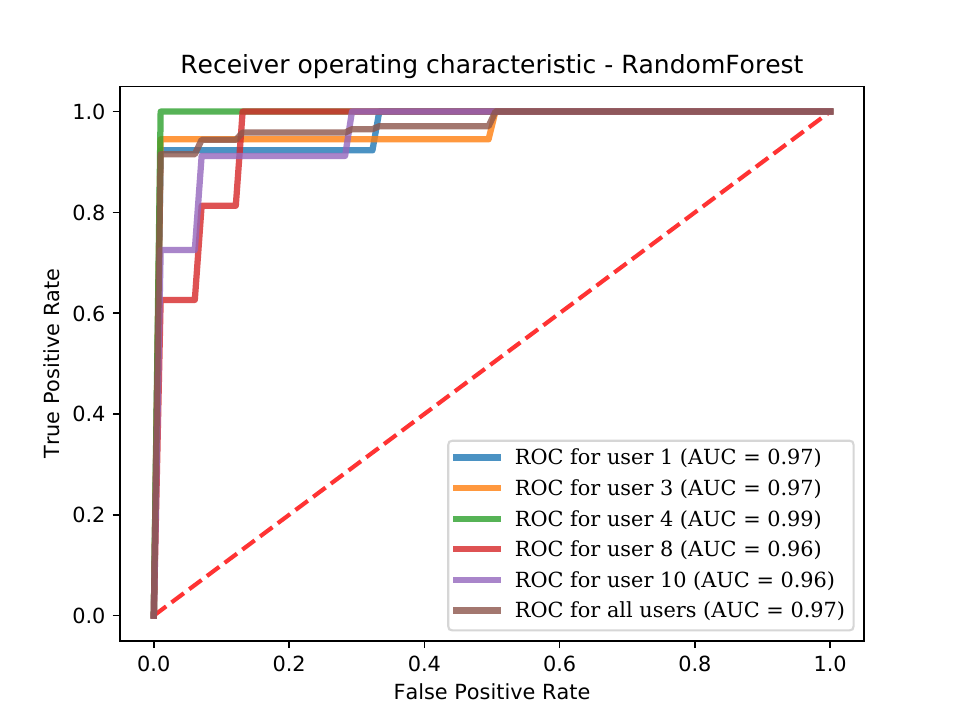}
		
	\end{minipage}
	%	\hspace{0.1cm}
	\begin{minipage}[b]{0.3\linewidth}
		\centering
		\includegraphics[width=\textwidth]{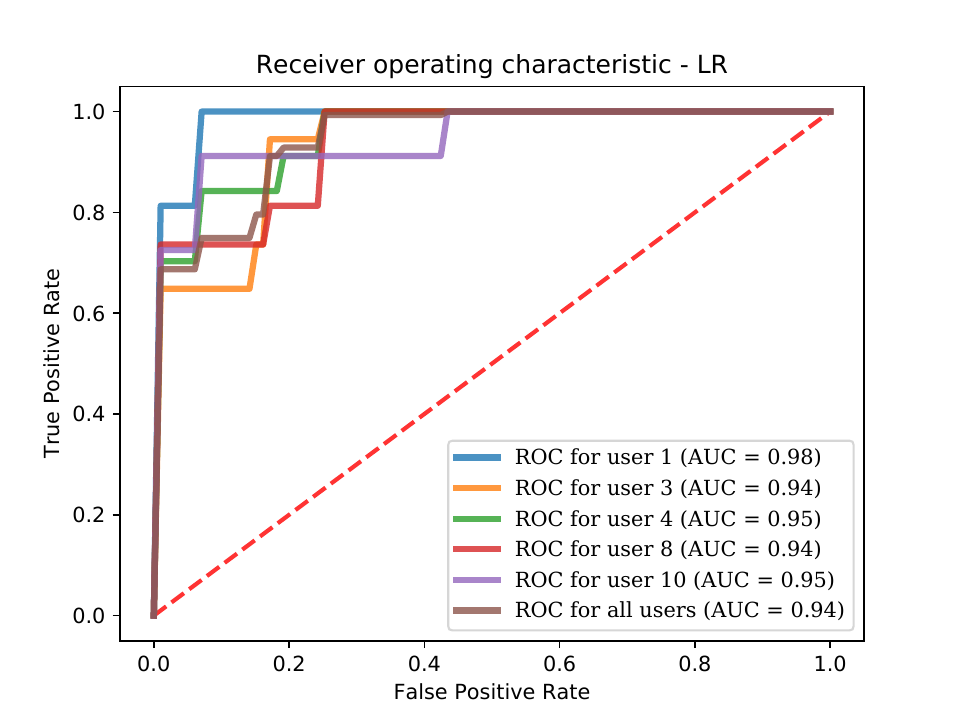}
		
	\end{minipage}
	%  \vspace{-1\baselineskip}
	%\captionsetup{justification=centering,margin=0cm}
	\caption{User-level ROC curves for 25 minute windows for three models among 6 users (top) and 5 users (bottom). Random Forest model outperforms the others in the one-vs-rest classification, and we observe difficulty of identifying each user separately.}
	\label{fig:figure_rocs5}

\end{figure*}

We compare three different models: Logistic Regression with $L_1$ regularization (LR), Random Forest (RF), and Gradient Boosting (GB) for time windows \added{with a length of} 25 minutes. We vary the hyper-parameters of these classifiers, and we select 2000 estimators for RF, and 2000 estimators for GB, with learning rate set at 0.01 and maximum depth at 3. For this experiment, we use the \emph{Device-only} features.  In Table~\ref{fig:figure_classifier_comparison}, we show user-level performance metrics for the three classifiers when all six users are considered. As observed, GB outperforms both RF and LR with its average precision and recall at 0.81 and 0.8, respectively. The performance of RF and GB is fairly close, but LR has lower classification performance (being a linear model with lower complexity). Additionally in Table~\ref{fig:figure_classifier_comparison}, we consider a setting with five users (removing User 6) and compare classification results. Surprisingly, by removing one of the users, we obtain a classification precision of 0.92 and recall of 0.92 with GB, an increase of more than 10\% compared to the setting with six users.  
We also generate ROC curves by training multiple binary-classifiers, each one for classifying one of the user classes versus the rest in Figure \ref{fig:figure_rocs5} for user groups. In this binary classification setting, Random Forest model outperforms the others, and we observe difficulty of identifying each user separately.

\subsection{High-Confidence Ensemble Model}

So far, we showed that models based on \emph{device-only} features perform relatively well at classifying users at time windows \added{with a length of} 25 minutes. As we discussed initially, different applications might have different requirements in terms of the confidence offered by the authentication score. For instance, a financial application might require high confidence in the user classification module before allowing a financial transaction. When the authentication score is used as a factor in a multi-factor authentication system, it could be acceptable to not compute an authentication score when the confidence is very low.
To account for these settings, we propose the idea of designing an ensemble of two models that are built independently and can be used in combination to increase the confidence in the authentication score. Our main insight is that we can build models leveraging the device-only and domain-only features independently, and compute an authentication score only when the two models agree on the user prediction. As a result, the overall confidence in the classification will increase. The cost is that, in situations when the two models disagree, no authentication score will be computed. In this case, the upper-level applications might wait for additional time, or leverage other authentication factors.

\begin{table}[t]

	\small
		\caption{Confusion matrix for the Gradient Boosting ensemble with one model using the device features, and the other the domain features, for 25 minute window for 6 users (top) and 5 users (bottom). If the classifiers disagree ("?"), no authentication score is computed. The classifier agrees ("Ag") most of the time, and high-confidence results improve compared to the single model approach. }
	\scalebox{0.8}{
			\renewcommand{\arraystretch}{1.2}
	\begin{tabular}{|c|c|c|c|c|c|c|c|c|c|c|c|}
		\hline
		& \textbf{U1} & \textbf{U3} & \textbf{U4} & \textbf{U6} & \textbf{U8} & \textbf{U10} & \textbf{?} & \textbf{Ag} & \textbf{Rec} & \textbf{Prec} & \textbf{F1}   \\ \hline
		\textbf{U1}     & 2          & 0          & 0          & 1          & 0          & 0           & 4                  & 3              & 0.66            & 0.66               & 0.66          \\ \hline
		\textbf{U3}     & 0          & 35         & 0          & 3          & 0          & 0           & 10                 & 38             & 0.92            & 0.92               & 0.92          \\ \hline
		\textbf{U4}     & 0          & 0          & 13         & 0          & 0          & 0           & 9                  & 13             & 1.0             & 1.0                & 1.0           \\ \hline
		\textbf{U6}     & 0          & 3          & 0          & 6          & 0          & 0           & 5                  & 9              & 0.66            & 0.54               & 0.6           \\ \hline
		\textbf{U8}     & 1          & 0          & 0          & 0          & 3          & 0           & 3                  & 4              & 0.75            & 0.75               & 0.75          \\ \hline
		\textbf{U10}    & 0          & 0          & 0          & 1          & 1          & 4           & 1                  & 6              & 0.66            & 1.0                & 0.8           \\ \hline
		\textbf{Total} & \multicolumn{6}{l|}{\textbf{}}                                               & \textbf{32}        & \textbf{73}    & \textbf{0.86}   & \textbf{0.87}      & \textbf{0.86} \\ \hline
		
	\end{tabular}}
%					%captionsetup{justification=centering,margin=2cm}
%					\caption{Confusion matrix for a Gradient Boosting ensemble with one model using the device features, and the other the domain features, for 25 minute window for 6 users. If the classifiers disagree, no authentication score is computed.\talha{explain symbols}}
%					\label{fig:multi_model_gb_gb_6user}
%\end{table}
%
%\begin{table}[t]
\newline
\newline

	\small
	\scalebox{0.8}{
			\renewcommand{\arraystretch}{1.2}
	\begin{tabular}{|c|c|c|c|c|c|c|c|c|c|c|}
		\hline
		
		\textbf{}      &\textbf{U1} & \textbf{U3} & \textbf{U4} & \textbf{U8} & \textbf{U10} & \textbf{?} & \textbf{Ag} & \textbf{Rec} & \textbf{Prec} & \textbf{F1}   \\ \hline
		\textbf{U1}     & 3          & 0          & 0          & 0          & 0           & 4                  & 3              & 1.0             & 1.0                & 1.0           \\ \hline
		\textbf{U3}     & 0          & 48         & 0          & 0          & 0           & 0                  & 48             & 1.0             & 0.97               & 0.98          \\ \hline
		\textbf{U4}     & 0          & 0          & 15         & 0          & 0           & 7                  & 15             & 1.0             & 1.0                & 1.0           \\ \hline
		\textbf{U8}     & 0          & 0          & 0          & 3          & 0           & 4                  & 3              & 1.0             & 0.75               & 0.85          \\ \hline
		\textbf{U10}    & 0          & 1          & 0          & 1          & 4           & 1                  & 6              & 0.66            & 1.0                & 0.8           \\ \hline
		\textbf{Total} & \multicolumn{5}{c|}{}                                           & \textbf{16}        & \textbf{75}    & \textbf{0.97}   & \textbf{0.97}      & \textbf{0.97} \\ \hline
	\end{tabular}}
	%	\vspace{1mm}
						%\captionsetup{justification=centering,margin=2cm}

	\label{fig:multi_model_gb_gb_5user}

\end{table}
We test the ensemble of two Gradient Boosting models, one built on the \emph{device-only} features, and the second on the \emph{domain-only} features.  The F1 score of the six-user model improves from 0.8 with a single GB model to 0.86 with the ensemble. In the ensemble, the models do not agree on 32 out of 105 sessions. For five users, the ensemble F1 score reaches 0.97 (compared to 0.92 for a single GB model). In this case, the ensemble does not compute a score on only 16 out of 91 sessions. We show the results of our ensemble user classification model in Table~\ref{fig:multi_model_gb_gb_5user}. Device and domain features capture different characteristics and helps the model have high-confidence if both models agree. The results demonstrate that the ensemble can reliably increase the confidence of the model, at the cost of not always providing an authentication score.

% !TEX root = iot_auth_main.tex
\section{Discussion}
We examined the feasibility of using behavioral authentication for a smart home using the correlation between user actions and network traffic. We conducted a user study in a realistic IoT environment and designed classifiers that  provide user authentication scores. The  scores can be used by various applications with different security policies. Below we discuss several challenges such an authentication system encounters in practice and directions for future work.

\myparagraph{Training ML system} We envision the use of traditional authentication methods (e.g. passwords or PINs on their cellphones) during the system training. Ground truth (labels) for training can be obtained from these traditional authentication methods. In a practical deployment, we envision that the ML model for user classification is periodically re-trained with new behavioral data over time. This is particularly important as new IoT devices are added to the smart home and users slowly shift their interests over time. During the system usage, in case of mis-classification, the user would be asked to provide a second factor authentication which could be used by the system for re-training. In terms of the ML model choice, boosting ensemble methods have the ability to assign higher weights to mis-classified points, and they could result in more robust models. For new users, training can be performed for some time period before the system starts generating authentication scores.\\
\myparagraph{Generating authentication scores} Our system generates authentication scores on a continuous basis, leveraging  information collected in the most recent observation period.  The scores can be used by the IoT devices residing at home (e.g. voice assistants, locks), as well as remote cloud applications to verify the transaction requested from user's home. We  emphasize that the system we propose is not meant to be used as the sole authentication mechanism for critical tasks, but it can enable more flexible multi-factor authentication policies customized per application. A financial application might require a high confidence score, in addition to the user's password, while changing the thermostat settings might rely solely on a medium-confidence authentication score.  We observed in our experimental evaluation that the authentication scores become more accurate and have higher confidence as the observation window increases. The reason is that in the smart home setting, users interact with the IoT devices intermittently.\\

\myparagraph{Applicability of neural networks} We also tested several deep neural network (DNN) architectures, including feed-forward neural networks and LSTMs. We did not include these details in the paper, as the accuracy of these models was not as strong as that of gradient boosting, random forest, and our ensemble.
The best F1 score for a feed-forward architecture was 0.54 and for LSTM 0.51, whereas our gradient boosting achieves an F1 score of 0.86. We suspect the reason is because we have a relatively limited dataset available for  training.  While in principle it is possible to conduct the user study over longer periods of time, we found it challenging to recruit and retain users over time. \\
\myparagraph{Impersonation attacks} Our system could in theory be vulnerable to user impersonation attacks when a strong attacker knows the exact actions of the targeted user and can inject traffic into the system. The adversary needs to spend some time in the smart home to be able to impact the score significantly, but this attack might be possible among people who share the living space. For short visitors or strangers, such an attack would be challenging to mount. On the other hand, remote adversaries might be able to send traffic on the network (through compromised mobile apps, for example). User impersonation by a remote adversary is difficult because the adversary cannot interact with all IoT devices remotely. However, a remote adversary might be able to inject some noise in the network traffic and cause mis-classification to a different user. This could result into a denial-of-service attack.\\

\myparagraph{Poisoning attacks} Poisoning has been demonstrated against a range of ML applications~\cite{biggio2012poisoning,Xiao15,Steinhardt17,Jagielski18}. In this setting, we suspect that an adversary needs access to the smart home  for longer periods of time to effectively poison the data collection. Additionally, the adversary's activities at training and testing time need to be carefully synchronized for the adversary to be successful to achieve his goals (for example, obtaining a high authentication score when the user is not in close proximity). We leave the investigation of the system's resilience to poisoning attacks as future work.

% !TEX root = iot_auth_main.tex

\section{Related work}
\label{sec:relwork}

Behavior-based authentication systems are designed to \added{identify} users based on their behavior. Implicit authentication \cite{shi2010implicit} shows the applicability of behavior modeling for authentication by utilizing the call/message information, browser activity, and GPS history in smart phones. Itus provides an extensible implicit authentication framework for Android~\cite{Itus}. A survey of multiple implicit authentication methods is given in \cite{SurveyImplicit}. Progressive Authentication \cite{riva2012progressive} models user behavior on mobile devices by combining biometric features and sensor data. Another emerging continuous authentication method leverages the sensor information from wearable devices (e.g. smart watches, activity trackers, glasses, bracelets) to learn user behavior~\cite{gafurov2006biometric,ZEBRA,peng2017continuous}.
Behavioral authentication in other contexts has also been studied. Freeman et al.~\cite{freeman2016you} designed an ML approach for clustering logins based on their IP, geolocation, browser, and time, for authenticating users to online services. Device fingerprinting has been used to augment user authentication on the web~\cite{Alaca16}.

Authorization and access policy frameworks in IoT systems has been an active research area. He et al.~\cite{he2018rethinking} conducted a survey of IoT device authorization preferences and observed that most devices have limited authorization flexibility. ContexIoT~\cite{ContexIoT} is a  permission system for IoT apps that provides users contextual information to enable them to make better access control decisions. SmartAuth~\cite{SmartAuth} generates a user interface for authorization decisions in IoT apps. Soteria~\cite{Soteria} performs static analysis to identify if IoT apps respect security policies. {\sf IoTGUARD}~\cite{IoTGuard} dynamically enforces security policies in IoT devices based on monitoring event handlers by IoT apps.	
There has been prior work investigating user authentication in the smart home environment. Apthorpe et al. \cite{apthorpe2017spying} show that user actions can be inferred from encrypted network traffic generated by IoT devices.   Emami-Naeini et al.~\cite{PrivacyIoT} perform a user study on IoT privacy and demonstrate that users are not comfortable with biometric data collection in IoT settings. Biometric-based authentication is optional in Alexa~\cite{alexa_voice_setting}, but this is prone to recording and replay attacks. Shi et al.~\cite{shi2017smart} proposed an authentication system leveraging physical properties of the WiFi signal generated by IoT devices. Krašovec, et al.~\cite{kravsovec2020not} propose a multi-modal approach for an IoT-based authentication system based on interactions with physical objects, combining it with data from the PC terminal and keyboard.

In this work, we utilize the network communication statistics of IoT devices through user interactions to build a user authentication framework using machine learning. Our system computes an authentication score for each user, which can be used by services to enforce flexible policies depending on the sensitivity of the service.

\section{Conclusion}
\label{sec:concl}

We propose a novel user authentication method by analyzing HTTPS network traffic generated by home-based IoT devices. We conduct a user study and experiment with several ML algorithms using features extracted from network traffic, with the goal of classifying users from a known set of users. We showed that random forest and gradient boosting models perform well in this setting and are able to obtain high accuracy at classifying users. Moreover, an ensemble of two models can increase the confidence of classification, at the expense of abstaining from generating a score when the models produce different predictions. The behavioral authentication scores computed by our authentication modules could be applied in various settings. More importantly, they open up the possibility of creating flexible policies for authorization and access control, and of replacing today's rigid, fixed policies. An interesting avenue for future research is multi-user authentication, i.e., design systems to learn the behavior of a group of users (e.g. a family) and provide authorization mechanisms to prevent an outsider from performing harmful actions, as well as detecting intrusions into the private living space. Studying the feasibility of adversarial impersonation and poisoning attacks in this setting is also of great interest. We believe that our work opens up new research avenues in these unexplored directions.  

\section*{Acknowledgements}

The authors would like to thank David Choffnes, Daniel Dubois, and Jingjing Ren for providing us access to the Mon(IoT)r Lab\footnote{https://moniotrlab.ccis.neu.edu} at Northeastern University and setting up the data collection infrastructure that enabled IoT device monitoring.  We thank Visa Research for funding this research. We would also like to thank Andres Molina-Markham for many discussions on contextual authentication.

\bibliographystyle{IEEEtran}

\bibliography{iot-auth}

\end{document}